\documentclass[preprint,showpacs,amsmath,amssymb,aps,prd,nofootinbib]{revtex4}
\usepackage{epsfig,graphicx,color,appendix}

\begin{document}
\begin{flushright}
KIAS-P13015
\end{flushright}

\title{\mbox{}\\[10pt]
Leptons and Quarks from a Discrete Flavor Symmetry}

\author{Y. H. Ahn\footnote{Email: yhahn@kias.re.kr}
}

\affiliation{School of Physics, KIAS, Seoul 130-722, Korea}

\date{\today}

\begin{abstract}
We propose a new model of leptons and quarks based on the discrete flavor symmetry $T'$, the double covering of $A_4$, in which the hierarchies of charged fermion masses and the mildness of neutrino masses are responsible for Higgs scalars. After spontaneous breaking of flavor symmetry, with the constraint of renormalizability in the Lagrangian, the leptons have $m_{e}=0$ and the quarks have the Cabibbo-Kobayashi-Maskawa (CKM) mixing angles $\theta^{q}_{12}=13^{\circ}, \theta^{q}_{23}=0^{\circ}$ and $\theta^{q}_{13}=0^{\circ}$. Thus, certain effective dimension-5 operators are introduced, which induce $m_{e}\neq0$ and lead the quark mixing matrix to the CKM one in form. On the other hand, the neutrino Lagrangian still keeps renormalizability. For completeness, we show numerical analysis: in the lepton sector, only normal mass hierarchy is permitted within $3\sigma$ experimental bounds with the prediction of both large deviations from maximality in the atmospheric mixing angle $\theta_{23}$ and the measured values of reactor angle. So, future precise measurements of $\theta_{23}$, whether $\theta_{23}\rightarrow45^{\circ}$ or $|\theta_{23}-45^{\circ}|\rightarrow5^{\circ}$, will either exclude or favor our model. Together with it, our model makes predictions for the Dirac CP phase, which is almost compatible with the global analysis in $1\sigma$ experimental bounds. Moreover, we show the effective mass $|m_{ee}|$ measurable in neutrinoless double beta decay to be in the range $0.04\lesssim|m_{ee}|[eV]<0.11$, which can be tested in near future neutrino experiments.
\end{abstract}

\maketitle %
\section{Introduction}

In the Standard Model (SM) of particle physics with a single Higgs there are enormously various hierarchies of quark and lepton Yukawa couplings, {\it that is}, $m_{t}/m_{u}=y_{t}/y_{u}\sim10^{5}, m_{\tau}/m_{e}=y_{\tau}/y_{e}\sim10^{3}$ etc.. In addition, there are several fundamental questions in Yukawa sector: why the top quark is uniquely so big compared with the other fermions, why neutrino masses are so small and so mild compared with the other charged fermions, why both the three leptonic mixing angles and one quark mixing angle are large, while the two quark mixing angles are so small. In some sense, our understanding of fermion masses and mixing angles remains at a very primitive level.
One of the approaches often adopted in the understanding for a possible solution for the flavor puzzle consists of the introduction of family symmetries which constrain the flavor structure of Yukawa couplings and lead to predictions for fermion masses and mixings. We propose a simplified way to address those questions in a non-Abelian discrete symmetry $T'$~\cite{Tprime}, by both introducing six types of Higgs fields and imposing all Yukawa couplings being of order one. Then the group $T'$ can be responsible for the present Pontecorvo-Maki-Nakagawa-Sakata (PMNS) and Cabibbo-Kobayashi-Maskawa (CKM) mixing matrices. And the hierarchies of fermions can originate from the different Higges.
The representations of $T'$ are those of $A_4$ plus three independent doublets ${\bf 2},{\bf 2}',{\bf 2}''$. Similar to $A_4$~\cite{A4,A4A, Ahn:2011yj,Ahn:2012cg, vacuum}~\footnote{The finite group $A_4$ describes the even permutations of four objects and possesses.}, in neutrino sector using four in-equivalent representations ${\bf 1}, {\bf 1}', {\bf 1}''$ and ${\bf 3}$ one can obtain the tri-bimaximal (TBM)~\cite{TBM} mixing pattern. In the presence of the doublet representations one can naturally describe the mass hierarchy among the charged fermions and the Cabibbo angle in the CKM matrix~\cite{PDG}.

After the relatively large reactor angle $\theta_{13}$ measured in Daya Bay and RENO~\cite{DayaRENO} including Double Chooz, T2K and MINOS experiments~\cite{Other}, the most recent analysis based on global fits~\cite{GonzalezGarcia:2012sz} of the neutrino oscillations enters into a new phase of precise determination of mixing angles and mass squared differences, indicating that the TBM mixing for three flavors should be corrected in the lepton sector:
their allowed ranges at $1\sigma$ $(3\sigma)$ from global fits are given by
 \begin{eqnarray}
  &&\theta_{13}=8.66^{\circ+0.44^{\circ}~(+1.30^{\circ})}_{~-0.46^{\circ}~(-1.47^{\circ})}~,\qquad\delta_{\rm CP}=300^{\circ+66^{\circ}~~(+60^{\circ})}_{~-138^{\circ}~(-300^{\circ})}~,\qquad\theta_{12}=33.36^{\circ+0.81^{\circ}~(+2.53^{\circ})}_{~-0.78^{\circ}~(-1.27^{\circ})}~,\nonumber\\
  &&\theta_{23}=40.0^{\circ+2.1^{\circ}}_{~-1.5^{\circ}}\oplus50.4^{\circ+1.3^{\circ}}_{~-1.3^{\circ}}~{1\sigma},\quad\left(35.8^{\circ}\thicksim54.8^{\circ}~{3\sigma}\right)~,\nonumber
 \end{eqnarray}
 \begin{eqnarray}
  &&\Delta m^{2}_{\rm Sol}[10^{-5}{\rm eV}^{2}]=7.50^{+0.18~(+0.59)}_{-0.19~(-0.50)}~,~\Delta m^{2}_{\rm Atm}[10^{-3}{\rm eV}^{2}]=\left\{\begin{array}{ll}
                2.473^{+0.070~(+0.222)}_{-0.067~(-0.197)}~, & \hbox{NMH} \\
                2.427^{+0.042~(+0.185)}_{-0.065~(-0.222)}~, & \hbox{IMH}~
                                  \end{array},
                                \right.
 \label{expnu}
 \end{eqnarray}
where $\Delta m^{2}_{\rm Sol}\equiv m^{2}_{2}-m^{2}_{1}$, $\Delta m^{2}_{\rm Atm}\equiv m^{2}_{3}-m^{2}_{1}$ for the normal mass hierarchy (NMH), and  $\Delta m^{2}_{\rm Atm}\equiv |m^{2}_{3}-m^{2}_{2}|$ for the inverted one (IMH).
While there are the large values of the solar mixing angle $\theta_{\rm sol}\equiv\theta_{12}$, the atmospheric mixing angle $\theta_{\rm atm}\equiv\theta_{23}$ and the reactor mixing angle $\theta_{\rm reac}\equiv\theta_{13}$ in the lepton sector, in the quark sector the Cabibbo angle and the other two small quark mixing angles, e.g., at $1\sigma$ level~\cite{ckmfitter} read:
 \begin{eqnarray}
  \theta^{q}_{12} = ( 13.03 \pm 0.05 )^{\circ}~,
  \quad\theta^{q}_{23} = ( 2.37^{+0.05}_{-0.09} )^{\circ}~,\quad\theta^{q}_{13} = ( 0.20^{+0.02}_{-0.02} )^{\circ} ~,\quad
  \delta^{q}_{CP} = ( 67.17_{-2.44}^{+2.78} )^{\circ} ~.
 \label{expquark}
 \end{eqnarray}
The discrepancy of the mixing angles in Eqs.~(\ref{expnu}) and (\ref{expquark}) may tell us about some new flavor symmetries of quarks and leptons. And it is well known that
the mass spectrum of the charged fermions exhibits a strong hierarchical pattern [see Eq.~(\ref{massRatio})], unlike that of neutrinos which shows a mild hierarchy. These facts may provide a clue to the nature of quark-lepton physics beyond the SM~\footnote{The dada (\ref{expnu}) and (\ref{expquark}) at $1\sigma$ seem to disfavor the maximal mixing in the atmospheric mixing angle, indicating that it starts to disfavor the Quark-Lepton Complementarity~\cite{QLC} on $\theta^{q}_{23}+\theta_{23}=45^{\circ}$, even it is not significant yet.}. Therefore, it is very important to find a natural model that leads to the observed flavor mixing patterns for quarks and leptons. In the present article we shall build such a model to emphasize the leptonic mixing parameters and the quarks one.

In this work we propose a new model based on flavor symmetry $T'$ that can accommodate quarks and leptons in the same framework invariant under $SU(2)_{L}\times U(1)_{Y}\times T'$. We introduce six types of Higgses in the Yukawa sector to depict
the mass hierarchies of fermions, and simultaneously the mixing parameters of the leptons and those of the quarks.
In addition, in order to simplify our model and to remove the unwanted Yukawa terms appearing in the Lagrangian, we impose a continuous global symmetry which cannot be gauged. After spontaneous $U(1)_{X}$ breaking, to avoid Goldstone bosons it has to be explicitly broken down to a subgroup.
We stress that in our model $CP$ invariance is originally explicitly broken in the Lagrangian level by the complex Clebsch-Gordan coefficients, even though all parameters in Lagrangian being real are imposed. The lepton sector in our model can not only naturally explain large deviations from the TBM but also provide a possibility for low-energy $CP$ violation in neutrino oscillations and the mildness of neutrino masses.


The paper is organized as follows. In the next section, we present the particle content together with the flavor symmetry of our model and the mass terms of both neutrino and charged fermion sectors after flavor symmetry breaking. In Sec.~$\textrm{III}$, we show the neutrino masses generated in a type seesaw-I and their mixing angles and $CP$ violation as well as the CKM matrix.  In Sec.~$\textrm{IV}$, for completeness, we show numerical analysis in the lepton sector.
Then, we give the conclusion in Sec.~$\textrm{IV}$, and briefly mention about the VEV alignments and spontaneous $CP$ violation, as an example, in Appendix.

\section{The Model}

In the absence of flavor symmetries, particle masses and mixings
are generally undetermined in a gauge theory. Here, we present a discrete symmetry model based on a $T'$ flavor symmetry for leptons and quarks in order to depict the mass hierarchies of charged fermions, the mildness of neutrino masses, and simultaneously the present mixing parameters
of the neutrino oscillation data and those of the quarks. 
Moreover, we describe the model to understand $CP$ violations in the lepton sector which is imperative, if the baryon asymmetry of the Universe (BAU) originated from leptogenesis scenario in the seesaw models~\cite{review}.

Here we recall that $T'$ is the symmetry group of the double tetrahedron \cite{Aranda:1999kc, Tprime}.
The group $T'$ has 24 elements and has two kinds of representations. It contains the representations of $A_4$: one triplet {\bf 3} and three singlets ${\bf 1},{\bf 1}'$ and ${\bf 1}''$. When working with these representations there is no distinction between the group $T'$ and the group $A_4$. In particular, in these representations, the elements of $T'$ coincide in pairs and can be described by the same matrices that represent the elements in $A_4$. The other representations are three boublets ${\bf 2},{\bf 2}'$ and ${\bf 2}''$. The representations ${\bf 1}',{\bf 1}''$ and ${\bf 2}',{\bf 2}''$ are complex conjugated to each other. Note that $A_4$ is not a subgroup of $T'$, since the two-dimensional representations can not be decomposed into representations of $A_4$.
The generators $S$ and $T$ satisfy the relation $S^{2}=R,T^{3}=(ST)^{3}=R^2={\bf 1}$, where $R={\bf 1}$ in case of the odd-dimensional representation and $R=-{\bf 1}$ for ${\bf 2},{\bf 2}',{\bf 2}''$ such that $R$ commutes with all elements of the group. In the three-dimensional unitary representation, there are abelian subgroups of $T'$ : $Z_{3}, Z_{4}$ and $Z_6$ symmetries, which are generated by the elements
 \begin{eqnarray}
T={\left(\begin{array}{ccc}
 1 &  0 &  0 \\
 0 &  \omega &  0 \\
 0 &  0 &  \omega^{2}
 \end{array}\right)}~,\qquad TST^{2}=\frac{1}{3}{\left(\begin{array}{ccc}
 -1 &  2 &  2 \\
 2 &  -1 & 2 \\
 2 & 2 &  -1
 \end{array}\right)}~,\qquad S=\frac{1}{3}{\left(\begin{array}{ccc}
 -1 &  2\omega &  2\omega^2 \\
 2\omega^2  &  -1 &  2\omega \\
 2\omega &  -1 &  2\omega^{2}
 \end{array}\right)}~,
 \label{generator}
 \end{eqnarray}
respectively, where $\omega=e^{i2\pi/3}$ is a complex cubic-root of unity. Especially, the elements $T$ and $TST^2$ are of importance for the structure of our model.
The group $T'$ has seven irreducible representations, one triplet ${\bf 3}$, three doublets ${\bf 2}, {\bf 2}', {\bf 2}''$ and three singlets ${\bf 1}, {\bf 1}', {\bf 1}''$
with the multiplication rules
${\bf 1}^{j}\otimes{\bf r}^{k}={\bf r}^{k}\otimes{\bf 1}^{j}={\bf r}^{j+k}$ for ${\bf r}=1,2$, ${\bf 1}^{j}\otimes{\bf 3}={\bf 3}\otimes{\bf 1}^{j}={\bf 3}$, ${\bf 2}^{j}\otimes{\bf 2}^{k}={\bf 3}\oplus{\bf 1}^{j+k}={\bf r}^{j+k}$, ${\bf 2}^{j}\otimes{\bf 3}={\bf 3}\otimes{\bf 2}^{j}={\bf 2}\oplus{\bf 2}'\oplus{\bf 2}''$ and
 ${\bf 3}\otimes{\bf 3}={\bf 3}_{s}\oplus{\bf 3}_{a}\oplus{\bf 1}\oplus{\bf 1}'\oplus{\bf 1}''$, where $j,k=0,\pm1$ and we have denoted ${\bf 1}^{0}\equiv{\bf 1}, {\bf 1}^{1}\equiv{\bf 1}', {\bf 1}^{-1}\equiv{\bf 1}'' $  and similarly for the doublet representations. The sum $j+k$ is modulo 3. The Clebsch-Gordan coefficients for the decomposition of the product representations are shown in Ref.~\cite{Feruglio:2007uu, Aranda:1999kc}.

We extend the standard model (SM) by the inclusion of an $T'$-triplet of
right-handed $SU(2)_{L}$-singlet Majorana neutrinos $N_{R}$, and the introduction of six types of scalar Higgs fields: the SM $SU(2)_{L}$-doublet Higgs bosons $\Phi,\Psi$, which we take to be $T'$-triplets $\mathbf{3}$ representation, the other two $SU(2)_L$-doublet Higgs bosons $H,G$, which are distinguished from $\Phi,\Psi$ by being  $T'$-doublets $\mathbf{2}$ representation, and another $SU(2)_{L}$-doublet $\mathbf{2}$ of Higgs boson $\eta$, which is a $T'$-singlet $\mathbf{1}$ representation, finally an $SU(2)_{L}$-singlet $T'$-triplet $\mathbf{3}$ Higgs field $\chi$:
 \begin{eqnarray}
  \Phi_j &=&
\begin{pmatrix} \varphi^{+}_j \\ \varphi^{0}_j \end{pmatrix},
\qquad~
H_k =
\begin{pmatrix}
  \phi^{+}_k \\
  \phi^{0}_k
\end{pmatrix} , \qquad \chi_j,\nonumber\\
 \Psi_j &=&
\begin{pmatrix}
  \psi^{+}_j \\
  \psi^{0}_j
\end{pmatrix},~\qquad
 G_k =
\begin{pmatrix} G^{+}_k \\ G^{0}_k \end{pmatrix},
\qquad
\eta =
\begin{pmatrix}
  \eta^{+} \\
  \eta^{0}
\end{pmatrix}
\label{Higgs}
 \end{eqnarray}
where $j=1,2,3$ and $k=1,2$. According to the above Higgs scalars, we impose $T'$ flavor symmetry for leptons and quarks. And, we levy all Yukawa couplings are of order one which implies that all the hierarchies of fermions appearing in the Lagrangian are responsible for the Higgses we introduced economically. In addition, after spontaneous breaking of flavor symmetry, the VEVs (vacuum expectation values) of such fields need spontaneous $CP$ violation. The representations of the field content of the model under $SU(2)\times U(1)\times T^{\prime}$ are summarized in Table-\ref{reps}, where each flavor of lepton doublets is assigned to one of the three $T'$-singlet representations: the electron-flavor to the ${\bf 1}$, the muon flavor to the ${\bf 1}''$, and the tau flavor to the ${\bf 1}'$, and $Q_{L}$ denotes left handed quark $SU(2)_L$ doublet and $\tau_{R}, \mathcal{E}_R$ and $t_{R}, \mathcal{U}_R$ $(b_R,\mathcal{D}_R)$ are the respective SM right handed lepton and $u$-type ($d$-type) $SU(2)_L$ singlets, respectively. Here, the down-type fermions $\mathcal{E}_R$, $\mathcal{D}_R$ and the up-type fermions $\mathcal{U}_R$ are assigned as $T'$-doublets and right handed gauge singlets:
 \begin{eqnarray}
  \mathcal{E}_R\left\{
                 \begin{array}{ll}
                   e_{R} \\
                   \mu_{R}
                 \end{array}
               \right.,\qquad
  \mathcal{D}_R\left\{
                 \begin{array}{ll}
                   d_{R} \\
                   s_{R}
                 \end{array}
               \right.,\qquad
  \mathcal{U}_R\left\{
                 \begin{array}{ll}
                   u_{R} \\
                   c_{R}
                 \end{array}
               \right.~.
 \end{eqnarray}

In the presence of three $T'$-triplet Higgs scalars $\Phi, \Psi$, $\chi$ and two $T'$-doublets Higgs scalars $H,G$, Higgs potential Lagrangian involving interaction terms among $\Phi$, $\chi$, $H, G$, $\Psi$ and $\chi$, which would be written as $V(\Phi\chi)$, $V(\Phi H), V(\Phi G), V(\Psi H), V(\Psi G)$, $V(H\chi), V(G\chi)$ and $V(\Psi\chi)$, would be problematic for vacuum stability. Such stability problems can be naturally solved, for instance, in the presence of extra dimensions or in supersymmetric dynamical completions~\cite{vacuum}. In these cases, those interaction terms are either disallowed or highly suppressed.
In our model, the $T'$ flavor symmetry is spontaneously broken by those $T'$-triplet and doublet scalars and $T'$-singlet scalar.
From the condition of the global minima of the scalar potential, we can obtain vacuum alignments of the fields $\chi,\Phi,\Psi, H, G$
relevant to achieve our goal. The Higgs potential of our model contains many terms, which is listed in Appendix~\ref{AHiggs}, Eqs.~(\ref{pot1})-(\ref{pot2}).
We spontaneously break the $T'$ flavor symmetry by giving non-zero vacuum expectation values to some components of the $T'$-triplets $\chi$, $\Psi$ and $\Phi$. We take the $T'$ symmetry breaking scale to be above the electroweak scale in our model, {\it i.e.}, $\langle\chi\rangle>\langle\Phi^{0}\rangle$.
As seen in Appendix~\ref{AHiggs}, the minimization of our scalar potential gives the following vacuum expectation values (VEVs):
 \begin{eqnarray}
\langle\chi\rangle&=&v_{\chi}e^{i\varphi}(1,1,1)~,\quad \qquad \langle\Phi^{0}\rangle=\frac{v_{\Phi}e^{i\gamma}}{\sqrt{2}}(1,0,0)~,\quad \qquad \langle\Psi^{0}\rangle=\frac{v_{\Psi}e^{i\zeta}}{\sqrt{2}}(1,0,0)~,\nonumber\\
\langle\eta^{0}\rangle&=& \frac{v_{\eta}}{\sqrt{2}}~,
\qquad~\langle H^{0}\rangle= \frac{1}{\sqrt{2}}(v_{H_1}e^{i\rho_1},v_{H_2}e^{i\rho_2})~,\qquad\langle G^{0}\rangle= \frac{1}{\sqrt{2}}(v_{G_1}e^{i\sigma_1},v_{G_2}e^{i\sigma_2})~.
 \label{vev}
 \end{eqnarray}
The SM VEV $v=(\sqrt{2}G_{F})^{-1/2}=246$ GeV results from the combination $v^{2}=\Sigma_{k}(v^{2}_{\eta}+v^{2}_{\Psi}+v^{2}_{H_k}+v^{2}_{G_k}+v^{2}_{\Phi})$ where $k=1,2$. The non-zero expectation value $\langle\chi\rangle\sim(1,1,1)$ breaks $T'$ symmetry down to a residual $Z_{4}$ symmetry which is generated by the group element $TST^2$.  The non-zero expectation values $ \langle\Phi^{0}\rangle\sim(1,0,0)$ and $\langle\Psi^{0}\rangle\sim(1,0,0)$ break $T'$ symmetry down to its subgroup $Z_{3}$ which is generated by the group element $T$. The non-zero expectation value $\langle\eta^{0}\rangle=v_{\eta}/\sqrt{2}$ does not break the $T'$ symmetry, because it is $T'$-flavorless. The non-zero expectation values $\langle H^{0}\rangle$ and $\langle G^{0}\rangle$
break $T'\rightarrow nothing$ with hierarchical breakings.
\begin{table}[t]
\begin{widetext}
\begin{center}
\caption{\label{reps} Representations of the fields under $T'$ and $SU(2)_L \times U(1)_Y$. }
\begin{ruledtabular}
\begin{tabular}{ccccccccccccc}
Field &$L_{\tau},L_{\mu},L_{e}$&$Q_{L}$&$\tau_R,\mathcal{E}_R$&$t_R,\mathcal{U}_R$&$b_R,\mathcal{D}_R$&$N_{R}$&$\chi$&$\eta$&$\Phi$&$H$&$\Psi$&$G$\\
\hline
$T'$&$\mathbf{1}'$, $\mathbf{1}''$, $\mathbf{1}$&$\mathbf{3}$ &$\mathbf{1}'$, $\mathbf{2}''$&$\mathbf{1}'$, $\mathbf{2}'$&$\mathbf{1}'$, $\mathbf{2}''$&$\mathbf{3}$&$\mathbf{3}$&$\mathbf{1}$&$\mathbf{3}$&$\mathbf{2}$&$\mathbf{3}$&$\mathbf{2}$\\
$SU(2)_L\times U(1)_Y$&$(2,-\frac{1}{2})$&$(2,\frac{1}{6})$&$(1,-1)$&$(2, \frac{2}{3})$&$(2,-\frac{1}{3})$&$(1,0)$&$(1,0)$&$(2,\frac{1}{2})$&$(2,\frac{1}{2})$&$(2,\frac{1}{2})$&$(2,\frac{1}{2})$&$(2,\frac{1}{2})$\\
\end{tabular}
\end{ruledtabular}
\end{center}
\end{widetext}
\end{table}
In addition to $T'$ flavor symmetry, we impose an additional symmetry $U(1)_{X}$ which is continuous global symmetry, where $L_{e}, L_{\mu}, L_{\tau}$, $\tau_{R}$, $\mathcal{E}_{R}$, $b_{R}$ and $\mathcal{D}_{R}$ carry $X=1$ and $\Phi$, $G$ carry $X=-1$, while all other fields have $X=0$. So this non-flavor symmetry forbids some irrelevant the $SU(2)_L\times U(1)_Y\times T^{\prime}$ invariant Yukawa terms from the Lagrangian (see later). And Since Goldstone bosons resulting from spontaneous $U(1)_X$ breaking via $\langle\Phi\rangle, \langle G\rangle\neq0$ are not allowed phenomenologically, so the additional symmetry $U(1)_{X}$ has to be explicitly broken~\footnote{In Appendix~\ref{AHiggs}, there are interaction terms $(\Phi^{\dag}\Psi)(\Phi^{\dag}\Psi), (G^{\dag}H)(G^{\dag}H)$ which break explicitly $U(1)_{X}$ to remove the unwanted Goldstone bosons in the low energy spectrum.} down to a subgroup $Z_{2}$ under which $\Phi\rightarrow-\Phi, G\rightarrow-G$ and the fields $L_{e}, L_{\mu}, L_{\tau}$, $\tau_{R}$, $\mathcal{E}_{R}$, $b_{R}$ and $\mathcal{D}_{R}$ also switch sign.

In our Lagrangian, we assume that there is a cutoff scale $\Lambda$, above which there exists unknown physics.
\subsection{The neutrino sector}
The Yukawa interactions ($d\leq5$) in the neutrino sector invariant under $SU(2)_L\times U(1)_Y\times T^{\prime}$ can be written as
 \begin{eqnarray}
 -{\cal L}^{\nu}_{\rm Yuk} &=& y^{\nu}_{1}\bar{L}_{\tau}(\tilde{\Phi}N_{R})_{{\bf 1}'}+y^{\nu}_{2}\bar{L}_{\mu}(\tilde{\Phi}N_{R})_{{\bf 1}''}+y^{\nu}_{3}\bar{L}_{e}(\tilde{\Phi}N_{R})_{{\bf 1}}\nonumber\\
 &+&\frac{1}{2}M(\overline{N^{c}_{R}}N_{R})_{{\bf 1}}+\frac{1}{2}y_R^\nu(\overline{N^{c}_{R}}N_{R})_{{\bf 3}_{s}}\chi+\text{h.c.},
 \label{lagrangiannu}
 \end{eqnarray}
where $\tilde{\Phi}\equiv i\tau_{2}\Phi^{\ast}$ and $\tau_{2}$ is a Pauli matrix. Note here that there are no dimension-5 operators driven by $\chi$ field in the neutrino sector, and the above Lagrangian in neutrino sector is renormalizable. In this Lagrangian, each flavor of neutrinos has its own independent Yukawa term, since they belong to different singlet representations ${\bf 1}'$, ${\bf 1}''$, and ${\bf 1}$ of $T'$: the neutrino Yukawa terms involve the $T'$-triplets $\Phi$ and $N_R$, which combine into the appropriate singlet representation. The right-handed neutrinos have an additional Yukawa term that involves the $T'$-triplet SM-singlet Higgs $\chi$.
The mass term $\frac{1}{2}M(\overline{N^{c}_{R}}N_{R})_{{\bf 1}}$ for the right-handed neutrinos is necessary to implement the seesaw mechanism by making the right-handed neutrino mass parameter $M$ large. The additional symmetry $U(1)_{R}$, as shown before, guarantees that the $SU(2)_L\times U(1)_Y\times T^{\prime}$ invariant Yukawa terms $\bar{L}_{e,\mu,\tau}\tilde{\Psi}N_{R}$ are forbidden from the Lagrangian.

After the breaking of the flavor and electroweak symmetries, with the VEV alignments as in Eq.~(\ref{vev}) the Dirac neutrino and right-handed neutrino mass terms from the Lagrangian (\ref{lagrangiannu}) result in
 \begin{eqnarray}
 -{\cal L}^{\nu}_{m} &=& \frac{v_{\Phi}e^{i\gamma}}{\sqrt{2}}\left(y^{\nu}_{3}\bar{\nu}_{e}N_{R1}+ y^{\nu}_{2}\bar{\nu}_{\mu}N_{R3}+y^{\nu}_{1}\bar{\nu}_{\tau}N_{R2}\right)\nonumber\\
&+&\frac{M}{2}(\overline{N^{c}_{R1}}N_{R1}+\overline{N^{c}_{R2}}N_{R3}+\overline{N^{c}_{R3}}N_{R2})
 +\frac{y_R^\nu v_{\chi}e^{i\varphi}}{6}\Big\{2\overline{N^{c}_{R1}}N_{R1}+2\overline{N^{c}_{R2}}N_{R2}
 +2\overline{N^{c}_{R3}}N_{R3}\nonumber\\
 &-&\overline{N^{c}_{R2}}N_{R3}-\overline{N^{c}_{R3}}N_{R2}-\overline{N^{c}_{R1}}N_{R2}-\overline{N^{c}_{R2}}N_{R1}-\overline{N^{c}_{R1}}N_{R3}-\overline{N^{c}_{R3}}N_{R1}\Big\}+\text{h.c.}~.
 \label{lagrangiannu1}
 \end{eqnarray}
Then, the neutrino Dirac mass terms and the right-handed Majorana neutrino mass terms are expressed as
\begin{align}
m_{D} & = \frac{v_{\Phi}e^{i\gamma}}{\sqrt{2}} {\left(\begin{array}{ccc}
 y^{\nu}_{3}  &  0 &  0 \\
 0 &  0 &  y^{\nu}_{2} \\
 0 &  y^{\nu}_{1} &  0 \\
 \end{array}\right)}
\label{eq:Ynu}
\\
 M_{R}&={\left(\begin{array}{ccc}
 M+\frac{2}{3}y^{\nu}_{R}\upsilon_{\chi}e^{i\varphi} &  -\frac{1}{3}y^{\nu}_{R}\upsilon_{\chi}e^{i\varphi} &  -\frac{1}{3}y^{\nu}_{R}\upsilon_{\chi}e^{i\varphi} \\
 -\frac{1}{3}y^{\nu}_{R}\upsilon_{\chi}e^{i\varphi} &  \frac{2}{3}y^{\nu}_{R}\upsilon_{\chi}e^{i\varphi} &  M-\frac{1}{3}y^{\nu}_{R}\upsilon_{\chi}e^{i\varphi} \\
 -\frac{1}{3}y^{\nu}_{R}\upsilon_{\chi}e^{i\varphi} &  M-\frac{1}{3}y^{\nu}_{R}\upsilon_{\chi}e^{i\varphi} &  \frac{2}{3}y^{\nu}_{R}\upsilon_{\chi}e^{i\varphi}
 \end{array}\right)}~,
 \label{MR}
\end{align}
where $v_{\Phi}$, $y^{\nu}_{1,2,3}$ and $M,y^{\nu}_{R},v_{\chi}$ are real positive variables.

\subsection{Charged fermion sector}
In the charged fermion sector, the Yukawa interactions ($d\leq5$) including dimension-5 operators driven by the $\chi$ field, invariant under $SU(2)_L\times U(1)_Y\times T^{\prime}$, are given by
 \begin{eqnarray}
 {\cal L}^{f}_{\rm Yuk} &=& {\cal L}^{d}_{\rm Yuk}+{\cal L}^{u}_{\rm Yuk}+{\cal L}^{\ell}_{\rm Yuk}~,
 \label{lagrangianCh}
 \end{eqnarray}
 where
 \begin{eqnarray}
 -{\cal L}^{d}_{\rm Yuk} &=& y_{b}(\bar{Q}_{L}\Psi)_{{\bf 1}''}b_{R}+Y_{d}\bar{Q}_{L}(H\mathcal{D}_{R})_{{\bf 3}}\nonumber\\
 &+&\frac{y^{a(s)}_b}{\Lambda}[(\bar{Q}_{L}\Psi)_{{\bf 3}}\chi]_{{\bf 1}''} b_{R}+\frac{Y^{d}_1}{\Lambda}\bar{Q}_{L}(H\mathcal{D}_{R})_{{\bf 1}''}\chi+\frac{Y^{a(s)}_d}{\Lambda}\bar{Q}_{L}(H\mathcal{D}_{R})_{{\bf 3}}\chi+\text{h.c.},\\
 -{\cal L}^{u}_{\rm Yuk} &=& y_{t}(\bar{Q}_{L}\tilde{\Phi})_{{\bf 1}''}t_{R}+Y_{u}\bar{Q}_{L}(\tilde{G}\mathcal{U}_{R})_{{\bf 3}}\nonumber\\
 &+&\frac{y^{a(s)}_t}{\Lambda}[(\bar{Q}_{L}\tilde{\Phi})_{{\bf 3}}\chi]_{{\bf 1}''} t_{R}+\frac{Y^{u}_1}{\Lambda}\bar{Q}_{L}(\tilde{G}\mathcal{U}_{R})_{{\bf 1}'}\chi+\frac{Y^{a(s)}_u}{\Lambda}\bar{Q}_{L}(\tilde{G}\mathcal{U}_{R})_{{\bf 3}}\chi+\text{h.c.}~,\\
 -{\cal L}^{\ell}_{\rm Yuk} &=& y_{\tau}\bar{L}_{\tau}\eta~ \tau_{R}+Y_{\mu}\bar{L}_{\mu}(H\mathcal{E}_R)_{{\bf 1}''}\nonumber\\
 &+&\frac{Y^{\mu}_1}{\Lambda}\bar{L}_{\mu}[(H\mathcal{E}_R)_{{\bf 3}}\chi]_{{\bf 1}''}+\frac{Y^{\mu}_2}{\Lambda}\bar{L}_{e}[(H\mathcal{E}_R)_{{\bf 3}}\chi]_{{\bf 1}}+\text{h.c.},
 \label{lagrangianCh1}
 \end{eqnarray}
with $\tilde{G}\equiv i\tau_{2}G^{\ast}$. In the charged-lepton sector tau lepton involves the $T'$-singlet $\eta$ and the $T'$-singlet right-handed charged-lepton $\tau_R$, while muon lepton has the $T'$-doublet $H$ and the $T'$-doublet right-handed charged-lepton $\mathcal{E}_{R}$ and there is no corresponding electron lepton term in renormalizable terms, indicating directly electron mass can be generated by the dimension-5 operators  driven by $\chi$ field. And the $\tau$-mass is generated upon the breaking of $T'-$invariant. Thus, the third family in charged leptons is different from the two, muon and electron.
On the other hand, in the quark sectors the use of $T'$-doublets $H,G$ and triplets $\Phi,\Psi$ Higgses can allow the third family to differ from the first two, and thus make plausible the mass hierarchies $m_{t}\gg m_{c}\gg m_{u}$ and $m_{b}\gg m_{s}\gg m_{d}$~\cite{PDG}. And in the renormalizable terms the $b$-quark and $t$-quark masses are generated upon the breaking of $T'\rightarrow Z_{3}$.
Mass terms of the quarks have two independent Yukawa terms with different couplings $(y_{b}, Y_{d})$ and $(y_{t}, Y_{u})$ for down-type quark and up-type quark, respectively, all involving $T'$-triplet Higgs fields $\Phi, \Psi$ and doublet fields $H, G$. The $T'$-triplet $\Phi$ is shared by both the three neutrino Yukawa terms and top-quark Yukawa term. The $T'$-doublet $H$ is involved by the terms associated to the $T'$-doublet right-handed lepton $\mathcal{E}_{R}$ and down-type quark $\mathcal{D}_{R}$. As mentioned before, the above Yukawa Lagrangian has the additional symmetry $U(1)_{X}$. This non-flavor symmetry (continuous Global symmetry) ensures that the $SU(2)\times U(1)\times T'$ invariant Yukawa terms $\bar{Q}_{L}(G\mathcal{D}_{R})_{{\bf 3}}$,$\bar{Q}_{L}(G\mathcal{E}_{R})_{{\bf 3}}$, $\bar{Q}_{L}(\tilde{H}\mathcal{U}_{R})_{{\bf 3}}$, $(\bar{Q}_{L}\Phi)_{{\bf 1}''}b_{R}$ and $(\bar{Q}_{L}\tilde{\Psi})_{{\bf 1}''}t_{R}$ are absent from the Lagrangian.

In the charged fermion sectors from the Lagrangian~(\ref{lagrangianCh}), after the breaking of the flavor and electroweak symmetries, with the VEV alignments as in Eq.~(\ref{vev}) the up-type quark and down-type quark mass terms result in
 \begin{eqnarray}
 {\cal L}^{f}_{m} &=&{\cal L}^{d}_{m}+{\cal L}^{u}_{m}+ {\cal L}^{\ell}_{m}~,
 \label{lagrangianQu1}
 \end{eqnarray}
where
 \begin{eqnarray}
-{\cal L}^{d}_{m} &=& y_{b}\frac{\tilde{v}_{\Psi}}{\sqrt{2}}\bar{b}_{L}b_{R}+ \frac{Y_{d}}{\sqrt{2}}\Big\{i\tilde{v}_{H_1}\bar{d}_{L}d_{R}+\frac{1-i}{2}\bar{s}_{L}\left(d_{R}\tilde{v}_{H_2}+s_{R}\tilde{v}_{H_1}\right)+\tilde{v}_{H_2}\bar{b}_{L}s_{R}\Big\}~\nonumber\\ &+&\frac{\tilde{v}_{\chi}}{\sqrt{2}\Lambda}\Big\{m^{d}_{33}~\bar{b}_{L}b_{R}+m^{d}_{23}~\bar{s}_{L}b_{R}+m^{d}_{13}~\bar{d}_{L}b_{R}+m^{d}_{12}~\bar{d}_{L}s_{R}+m^{d}_{22}~\bar{s}_{L}s_{R}+m^{d}_{32}~\bar{b}_{L}s_{R}\nonumber\\
&+&m^{d}_{11}~\bar{d}_{L}d_{R}+m^{d}_{21}~\bar{s}_{L}d_{R}+m^{d}_{31}~\bar{b}_{L}d_{R}\Big\}
+\text{h.c.}~,\label{lagrangianDown}\\
-{\cal L}^{u}_{m} &=& y_{t}\frac{\tilde{v}_{\Phi}}{\sqrt{2}}\bar{t}_{L}t_{R}
 +\frac{Y_{u}}{\sqrt{2}}\Big\{\tilde{v}_{G_2}\bar{u}_{L}c_{R}+i\tilde{v}_{G_1}\bar{c}_{L}u_{R}+\frac{1-i}{2}\bar{t}_{L}\left(u_{R}\tilde{v}_{G_2}+c_{R}\tilde{v}_{G_1}\right)\Big\}\nonumber\\
&+&\frac{\tilde{v}_{\chi}}{\sqrt{2}\Lambda}\Big\{m^{t}_{33}~\bar{t}_{L}t_{R}+m^{t}_{23}~\bar{c}_{L}t_{R}+m^{t}_{13}~\bar{u}_{L}t_{R}+m^{t}_{12}~\bar{u}_{L}c_{R}+m^{t}_{22}~\bar{c}_{L}c_{R}+m^{t}_{32}~\bar{t}_{L}c_{R}\nonumber\\
&+&m^{t}_{11}~\bar{u}_{L}u_{R}+m^{t}_{21}~\bar{c}_{L}u_{R}+m^{t}_{31}~\bar{t}_{L}u_{R}\Big\}
+\text{h.c.}~,\label{lagrangianUP}\\
-{\cal L}^{\ell}_{m} &=& y_{\tau}\frac{v_{\eta}}{\sqrt{2}}\bar{\tau}_{L}\tau_{R}
+\frac{Y_{\mu}}{\sqrt{2}}\bar{\mu}_{L}\left(e_{R}\tilde{v}_{H_2}-\mu_{R}\tilde{v}_{H_1}\right)\nonumber\\
&+&\frac{\tilde{v}_{\chi}}{\sqrt{2}\Lambda}\Big\{m^{\ell}_{12}~\bar{e}_{L}\mu_{R}+m^{\ell}_{22}~\bar{\mu}_{L}\mu_{R}+m^{\ell}_{11}~\bar{e}_{L}e_{R}+m^{\ell}_{21}~\bar{\mu}_{L}e_{R}\Big\}+\text{h.c.}~,
\label{lagrangianChLep}
 \end{eqnarray}
with $\tilde{v}_{\Psi}=v_{\Psi}e^{i\zeta}, \tilde{v}_{\Phi}=v_{\Phi}e^{i\gamma}, \tilde{v}_{G_k}=v_{G_k}e^{i\sigma_k}$, $\tilde{v}_{H_k}=v_{H_k}e^{i\rho_k}$ ($k=1,2$) and $\tilde{v}_{\chi}=v_{\chi}e^{i\varphi}$. In the above Eqs.~(\ref{lagrangianDown}), (\ref{lagrangianUP}) and (\ref{lagrangianChLep}), the entries $m^{d}_{ij}, m^{t}_{ij}$ and $m^{\ell}_{ij}$ are given in Appendix~\ref{sectionC}.
Then, the down-type quark mass matrix $\mathcal{M}_{d}$ is given by
  \begin{eqnarray}
 \mathcal{M}_{d}&=&\frac{1}{\sqrt{2}}{\left(\begin{array}{ccc}
 iY_{d}\tilde{v}_{H_1} & 0  &  0 \\
 \frac{1-i}{2}Y_{d}\tilde{v}_{H_2} &  \frac{1-i}{2}Y_{d}\tilde{v}_{H_1} &  0 \\
 0 &  Y_{d}\tilde{v}_{H_2} &  y_{b}\tilde{v}_{\Psi}
 \end{array}\right)}
 +\frac{v_{\chi}e^{i\varphi}}{\sqrt{2}\Lambda}{\left(\begin{array}{ccc}
 m^{d}_{11} & m^{d}_{12} &  m^{d}_{13} \\
 m^{d}_{21} & m^{d}_{22} &  m^{d}_{23} \\
 m^{d}_{31} & m^{d}_{32} &  m^{d}_{33}
 \end{array}\right)}\nonumber\\
 &=&V^{d}_{L}{\rm Diag}(m_{d},m_{s},m_{b})V^{d\dag}_{R}~.
 \label{dType}
 \end{eqnarray}
And, the up-type quark mass matrix $\mathcal{M}_{u}$ can be explicitly expressed as
 \begin{eqnarray}
 \mathcal{M}_{u}&=&\frac{1}{\sqrt{2}}{\left(\begin{array}{ccc}
 0 & Y_{u}\tilde{v}_{G_2} & 0 \\
 iY_{u}\tilde{v}_{G_1} & 0 &  0 \\
 \frac{1-i}{2}Y_{u}\tilde{v}_{G_2} &  \frac{1-i}{2}Y_{u}\tilde{v}_{G_1} &  y_{t}\tilde{v}_{\Phi}
 \end{array}\right)}
 +\frac{v_{\chi}e^{i\varphi}}{\sqrt{2}\Lambda}{\left(\begin{array}{ccc}
 m^{t}_{11} & m^{t}_{12} &  m^{t}_{13} \\
 m^{t}_{21} & m^{t}_{22} &  m^{t}_{23} \\
 m^{t}_{31} & m^{t}_{32} &  m^{t}_{33}
 \end{array}\right)}\nonumber\\
 &=&V^{u}_{L}{\rm Diag}(m_{u},m_{c},m_{t})V^{u\dag}_{R}~.
 \label{uType}
 \end{eqnarray}
Finally, with the VEV alignment in Eq.~(\ref{vev})
the charged-lepton mass matrix $\mathcal{M}_{\ell}$ can be explicitly expressed as
 \begin{eqnarray}
 \mathcal{M}_{\ell}&=&\frac{1}{\sqrt{2}}{\left(\begin{array}{ccc}
 0 & 0 & 0 \\
 -Y_{\mu}\tilde{v}_{H_2} & Y_{\mu}\tilde{v}_{H_1} &  0 \\
 0 &  0 &  y_{\tau}v_{\eta}
 \end{array}\right)} +\frac{v_{\chi}e^{i\varphi}}{\sqrt{2}\Lambda}{\left(\begin{array}{ccc}
 m^{\ell}_{11} & m^{\ell}_{12} &  0 \\
 m^{\ell}_{21} & m^{\ell}_{22} &  0 \\
 0 & 0 &  0
 \end{array}\right)}\nonumber\\
 &=&V^{\ell}_{L}~{\rm Diag}(m_{e},m_{\mu},m_{\tau})~V^{\ell\dag}_{R}~.
 \label{charged-lepton}
 \end{eqnarray}
In Eqs.~(\ref{dType}), (\ref{uType}) and (\ref{charged-lepton}), $V^{f}_{L}$ and $V^{f}_{R}$ are the diagonalization matrices for $\mathcal{M}_{f}$.

There exist several empirical fermion mass hierarchies in the up- and down-type quark and charged-lepton sectors
calculated from the measured values~\cite{PDG} :
 \begin{eqnarray}
 m_{u}=2.4~{\rm MeV}\qquad\qquad\qquad\qquad\qquad m_{c}&=&1.27~{\rm GeV}\qquad m_{t}=171.2~{\rm GeV}\nonumber\\
 m_{d}=4.8~{\rm MeV}\quad m_{s}=104~{\rm MeV}\qquad~m_{b}&=&4.2~{\rm GeV}\nonumber\\
 m_{e}=0.511~{\rm MeV}\quad\qquad~~ m_{\mu}=105.7~{\rm MeV}~\quad m_{\tau}&=&1.777~{\rm GeV}
 \label{massRatio}
 \end{eqnarray}
which implies that the possible quark-lepton symmetry~\cite{Pati:1973rp} is broken by the masses of quarks
and leptons. Thus, it is not expected that the known quark mixing pattern is transmitted to the lepton
sector in the exactly same form. In addition, a key point inferred from Eq.~(\ref{massRatio}) is
that the mass spectrum of the charged leptons exhibits a similar hierarchical pattern to that of the
down-type quarks except for the electron mass which is much smaller than $d$-quark one, unlike that of the up-type quarks which shows a much stronger hierarchical pattern and top-quark is uniquely biggest. Further, there is another interesting empirical relation
 \begin{eqnarray}
  |V_{us}| \approx \left( \frac{m_{d}}{m_{s}} \right)^{\frac{1}{2}}
  \approx 3 \left(\frac{m_{e}}{m_{\mu}} \right)^{\frac{1}{2}}~,
 \label{Vus}
 \end{eqnarray}
which has been known for quite a long time~\cite{Gatto:1968ss}.
For instance,  in terms of the Cabbibo angle $\lambda \equiv \sin\theta_{\rm C} \approx |V_{us}|$, the
fermion masses are scaled as ~$(m_{e},m_{\mu}) \approx (\lambda^{5},\lambda^{2})~ m_{\tau}$,
$(m_{d},m_{s}) \approx (\lambda^{4},\lambda^{2})~ m_{b}$~ and
~$(m_{u},m_{c}) \approx (\lambda^{8},\lambda^{4})~ m_{t}$,~ which may represent the followings: (i) there is at least one Higgs scalar shared by both charged-lepton and down-type quark sectors, or (ii) the mixing matrix of the charged lepton sector is similar to that of the down-type quark sector, and (iii) the CKM matrix is mainly generated by the mixing matrix of the down-type quark sector.

One of most interesting features observed by experiments on the charged fermions is that the mass spectra of quarks and charged leptons are strongly hierarchical, {\it i.e.}, the masses of third generation fermions are much heavier than those of the first and second generation fermions.
For the elements of $\mathcal{M}_{f}$ given in Eqs.~(\ref{uType}) and (\ref{dType}), taking into account the most natural case that the charged fermion masses have the strong hierarchy $m_{t}\gg m_{c}\gg m_{u}$ ($m_{b(\tau)}\gg m_{s(\mu)}\gg m_{d(e)}$) as well as Eq.~(\ref{massRatio}), we make a plausible assumption
 \begin{eqnarray}
 y_{t}v_{\Phi}\gg y_{b}v_{\Psi}=y_{\tau}v_{\eta}=Y_{u}v_{G_1}\gg Y_{d}v_{H_2}\gg Y_{u}v_{G_2}=Y_{d}v_{H_1}~.
 \label{ass}
 \end{eqnarray}
Then $V^{f}_{L}$ and $V^{f}_{R}$ can be determined by diagonalizing the matrices $\mathcal{M}_{f}\mathcal{M}^{\dag}_{f}$ and $\mathcal{M}^{\dag}_{f}\mathcal{M}_{f}$, respectively, indicated from Eqs.~(\ref{charged-lepton}), (\ref{uType}) and (\ref{dType}). Especially, the mixing matrix $V^{f}_{L}$ becomes one of the matrices composing the PMNS and CKM ones and it will be shown later. A general diagonalizing matrix $V^{f}_{L}$ can be parameterized in terms of three mixing angles and six phases:
 \begin{eqnarray}
 V^{f}_{L}={\left(\begin{array}{ccc}
 c^{f}_{2}c^{f}_{3} &  c^{f}_{2}s^{f}_{3}e^{i\phi^{f}_{3}} &  s^{f}_{2}e^{i\phi^{f}_{2}} \\
 -c^{f}_{1}s^{f}_{3}e^{-i\phi^{f}_{3}}-s^{f}_{1}s^{f}_{2}c^{f}_{3}e^{i(\phi^{f}_{1}-\phi^{f}_{2})} &  c^{f}_{1}c^{f}_{3}-s^{f}_{1}s^{f}_{2}s^{f}_{3}e^{i(\phi^{f}_{1}-\phi^{f}_{2}+\phi^{f}_{3})} &  s^{f}_{1}c^{f}_{2}e^{i\phi^{f}_{1}} \\
 s^{f}_{1}s^{f}_{3}e^{-i(\phi^{f}_{1}+\phi^{f}_{3})}-c^{f}_{1}s^{f}_{2}c^{f}_{3}e^{-i\phi^{f}_{2}} &  -s^{f}_{1}c^{f}_{3}e^{-i\phi^{f}_{1}}-c^{f}_{1}s^{f}_{2}s^{f}_{3}e^{i(\phi^{f}_{3}-\phi^{f}_{2})} &  c^{f}_{1}c^{f}_{2}
 \end{array}\right)}P_{f}~,
 \label{Vl}
 \end{eqnarray}
where $s^{f}_{i}\equiv \sin\theta^{f}_{i}$, $c^{f}_{i}\equiv \cos\theta^{f}_{i}$ and a diagonal phase matrix $P_{f}={\rm diag}(e^{i\xi^{f}_{1}},e^{i\xi^{f}_{2}},e^{i\xi^{f}_{3}})$ which can be rotated away by the phase redefinition of left-charged fermion fields.

\subsubsection{The down-type quark sector and its mixing matrix}
First, we consider the down-type quark sector.
From Eq.~(\ref{dType}) we see that the down-type quark mass matrix $\mathcal{M}_{d}$ can be diagonalized in the mass basis by a biunitary transformation,
$V^{d\dag}_{L}\mathcal{M}_{d}V^{d(\ell)}_{R}={\rm Diag}(m_{d},m_{s},m_{b})$.
The matrices $V^{d}_{L}$ and $V^{d}_{R}$ can be determined by diagonalizing the matrices
$\mathcal{M}_{d}\mathcal{M}^{\dag}_{d}$ and $\mathcal{M}^{\dag}_{d}\mathcal{M}_{d}$, respectively.
Especially, the left-handed down-type quark mixing matrix $V^{d}_{L}$ becomes one of the matrices
composing the CKM matrix such as $V_{\rm CKM} \equiv V^{u\dag}_{L}V^{d}_{L}$ [see Eq.~(\ref{CKM})].
From Eq.~(\ref{dType}) the hermitian square of the mass matrix for down-type quark $\mathcal{M}_{d}\mathcal{M}^{\dag}_{d}$ can be obtained. And, from Eqs.~(\ref{dType}) and (\ref{Vl}) the mixing angles and phases can be expressed in terms of Eq.~(\ref{ass}) as
 \begin{eqnarray}
 &&\theta^{d}_{1}\simeq\frac{v_{\chi}}{\Lambda}\left(\frac{y^{a}_{b}}{2y_{b}}+\frac{y^{s}_{b}}{3y_{b}}\right)~,\quad\qquad\theta^{d}_{2}\simeq\frac{v_{\chi}}{\Lambda}\frac{2y^{s}_{b}}{3y_{b}}~,\quad\qquad\theta^{d}_{3}\simeq\sqrt{2}\frac{v_{H_1}}{v_{H_2}}~,\nonumber\\
 &&~~\phi^{d}_{1}\simeq\frac{\varphi}{2}+\frac{\pi}{2}~,\quad\qquad\qquad\qquad\phi^{d}_{2}\simeq\frac{\varphi}{2}~,\quad\qquad\phi^{d}_{3}\simeq\frac{\rho_{12}}{2}+\frac{1}{2}\arg\left(i-1\right)~,
 \label{downMixing}
 \end{eqnarray}
where the parameters $y^{s}_{b}$ and $y^{a}_{b}$ are positive real numbers of order unity.
The empirical relation Eq.~(\ref{massRatio}) for down-type quark can be satisfied by setting as follows
 \begin{eqnarray}
 \frac{v_{\chi}}{\Lambda}\left(\frac{y^{a}_{b}}{2y_{b}}+\frac{y^{s}_{b}}{3y_{b}}\right)\equiv A_{d}\lambda^{2}~,\quad\qquad \frac{v_{\chi}}{\Lambda}\frac{2y^{s}_{b}}{3y_{b}}\equiv B_{d}\lambda^{3}~, \quad\qquad~v_{H_1}=\frac{\lambda}{\sqrt{2}} v_{H_2}~,
 \label{downMixing2}
 \end{eqnarray}
where the parameters $A_{d}$ and $B_{d}$ are positive real number of order unity. Note that the third relation in Eq.~(\ref{downMixing2}) comes from the renormalizable terms.
Then, from the above relation the mass squared eigenvalues are written as
 \begin{eqnarray}
 &&V^{d\dag}_{L}\mathcal{M}_{d}\mathcal{M}^{\dag}_{d}V^{d}_{L} \equiv{\rm Diag}\left(m^{2}_{d},~m^{2}_{s},~m^{2}_{b}\right)\nonumber\\
 &&\simeq{\rm Diag}\left(\frac{\lambda v^{2}_{H_2}}{\sqrt{2}}Y^{2}_{d}\frac{v_{\chi}}{\Lambda}\left\{Y_1\sin(\rho_{12}-\varphi)+ Y_{2}\cos(\rho_{12}-\varphi)\right\},~\frac{1}{4}Y^{2}_{d}v^{2}_{H_2},~ \frac{1}{2}y^{2}_{b}v^{2}_{\Psi}\right)~,
 \label{downMass}
 \end{eqnarray}
where $Y_{d}v_{H_2}=\sqrt{2}\lambda^{2}y_{b}v_{\Psi}$, and $Y_1=\frac{Y^{d}_1}{Y_{d}}+\frac{Y^{a}_d}{4Y_{d}}+\frac{Y^{s}_d}{6Y_{d}}$, $Y_2=\frac{Y^{a}_d}{4Y_{d}}+\frac{Y^{s}_d}{6Y_{d}}$. So, one can obtain the measured value of $m_{d}/m_{b}\simeq\lambda^{4}$.
Then, we can obtain the mixing matrix $V^{d}_{L}$ of the down-type quarks: under the constraint of unitarity up to ${\cal O}(\lambda^{3})$, it can be written as
 \begin{eqnarray}
 V^{d}_{L}=
 {\left(\begin{array}{ccc}
 1 -\frac{\lambda^{2}}{2} & \lambda e^{i\phi^{d}_{3}}  & B_d \lambda^{3}e^{i\frac{\varphi}{2}}  \\
 - \lambda e^{-i\phi^{d}_{3}} &  1-\frac{\lambda^{2}}{2} & A_d \lambda^{2}e^{i(\frac{\varphi}{2}+\frac{\pi}{2})} \\
  A_d \lambda^{3} e^{-i(\phi^{d}_{3}+\frac{\varphi}{2}+\frac{\pi}{2})}-B_d \lambda^{3}e^{-i\frac{\varphi}{2}}
  &  -A_d \lambda^{2}e^{-i(\frac{\varphi}{2}+\frac{\pi}{2})} & 1
 \end{array}\right)}P_{d}+{\cal O}(\lambda^{4})~.
 \label{DL}
 \end{eqnarray}

\subsubsection{The up-type quark sector and its mixing matrix}
Next, let us consider the up-type quark sector to obtain the realistic CKM matrix. From Eq.~(\ref{uType}) and Eq.~(\ref{ass}) the hermitian square of the mass matrix for up-type quark $\mathcal{M}_{u}\mathcal{M}^{\dag}_{u}$, with the condition given in Eq.~(\ref{ass})
the mass squared eigenvalues are written in a good approximation as
 \begin{eqnarray}
  m^{2}_{t}&\simeq&\frac{1}{2}y^{2}_{t}v^{2}_{\Phi}~, \qquad m^{2}_{c}\simeq \frac{1}{2}Y^{2}_{u}v^{2}_{G_1}~,\qquad m^{2}_{u}\simeq \frac{1}{2}Y^{2}_{u}v^{2}_{G_2}~.
 \label{up-type mass}
  \end{eqnarray}
Then, one can set $m_u/m_t\simeq v_{G_2}/v_{\Phi}\simeq\lambda^{8}$ and $m_c/m_t\simeq v_{G_1}/v_{\Phi}\simeq\lambda^{4}$ for equal amounts of Yukawa couplings.
And, the mixing angles and phases can be expressed in terms of Eq.~(\ref{ass}) as
 \begin{eqnarray}
  &&\theta^{u}_{1}\simeq\frac{v_{\chi}}{\Lambda}\left(\frac{y^{a}_{t}}{2y_{t}}+\frac{y^{s}_{t}}{3y_{t}}\right)~, \qquad \theta^{u}_{2}\simeq\frac{v_{\chi}}{\Lambda}\frac{2y^{s}_{t}}{3y_{t}}~, \qquad  \theta^{u}_{3}\simeq\frac{v_{\chi}}{\Lambda}\left(\frac{Y^{a}_{u}}{2Y_{u}}+\frac{Y^{s}_{u}}{3Y_{u}}\right)~\nonumber\\
  &&~\quad\phi^{u}_{1}\simeq\frac{\varphi}{2}+\frac{\pi}{2}~, \qquad\qquad\qquad \phi^{u}_{2}\simeq\frac{\varphi}{2}~, ~\quad\qquad\qquad \phi^{u}_{3}\simeq\frac{\pi}{2}+\frac{\varphi}{2}~.
 \label{up-type Angle}
 \end{eqnarray}
Due to $v_{\chi}/\Lambda\sim\lambda^{2}$ in Eq.~(\ref{downMixing2}) and the measured value of $m_{u}/m_{c}\approx v_{G_2}/v_{G_1}\approx\lambda^{4}$ in Eq.~(\ref{massRatio}), it is impossible to generate the Cabbibo angle, $\lambda\approx|V_{us}|$, from the mixing between the first and second generations in the up-type quark sector: if one sets $|(V^{u}_{L})_{12}|=\theta^{u}_{3}$, then from Eqs.~(\ref{up-type mass}) and (\ref{up-type Angle}) one obtains $|(V^{u}_{L})_{12}|\sim v_{\chi}/\Lambda\approx\lambda^{2}$, in discrepancy with the measured $\lambda\approx|V_{us}|$.  And similar to Eq.~(\ref{downMixing2}), from Eq.~(\ref{up-type Angle}) one can set
 \begin{eqnarray}
  \theta^{u}_{1}&\simeq& A_{u}\lambda^{2}~, \qquad \theta^{u}_{2}\simeq B_{u}\lambda^{3}~, \qquad \theta^{u}_{3}\simeq C_{u} \lambda^{2}~,
 \label{up-type Angle1}
 \end{eqnarray}
where $A_{u},B_{u},C_{u}$ are positive real numbers of order unity. Then, plugging Eqs.~(\ref{up-type Angle}) and (\ref{up-type Angle1}) into Eq.~(\ref{Vl}) the up-type quark mixing matrix $V^{u}_{L}$ can be written, under the constraint of unitarity up to ${\cal O}(\lambda^{3})$, as
 \begin{eqnarray}
 V^{u}_{L}=
 {\left(\begin{array}{ccc}
 1 & C_u\lambda^{2}e^{i\frac{\varphi+\pi}{2}}  &  B_u\lambda^{3}e^{i\frac{\varphi}{2}}  \\
 -C_u \lambda^{2}e^{-i\frac{\varphi+\pi}{2}} & 1 & A_u \lambda^{2}e^{i(\frac{\varphi}{2}+\frac{\pi}{2})} \\
 -B_u \lambda^{3}e^{-i\frac{\varphi}{2}}
  &  -A_u \lambda^{2}e^{-i(\frac{\varphi}{2}+\frac{\pi}{2})} & 1
 \end{array}\right)}P_{u}+{\cal O}(\lambda^{4})~,
 \label{UL}
 \end{eqnarray}
which indicates that the mixing matrix in the up-type quark sector can affect, at most, the next leading order contributions in $\lambda$. Flavor-changing neutral currents (FCNCs) in up-type quarks may restrict $A_{u}, B_{u}, C_{u}<\lambda$, which we will not discuss.

\subsubsection{The charged lepton sector and its mixing matrix}
Similar to the quark sector, from Eq.~(\ref{charged-lepton}) we see that the charged lepton mass matrix $\mathcal{M}_{\ell}$ can be diagonalized in the mass basis by a biunitary transformation,
$V^{\ell\dag}_{L}\mathcal{M}_{\ell}V^{d(\ell)}_{R}={\rm Diag}(m_{e},m_{\mu},m_{\tau})$.
The matrices $V^{\ell}_{L}$ and $V^{\ell}_{R}$ can be determined by diagonalizing the matrices
$\mathcal{M}_{\ell}\mathcal{M}^{\dag}_{\ell}$ and $\mathcal{M}^{\dag}_{\ell}\mathcal{M}_{\ell}$, respectively.
Especially, the charged lepton mixing matrix $V^{\ell}_{L}$ becomes one of the matrices
composing the PMNS matrix such as $U_{\rm PMNS}\equiv V^{\ell\dag}_{L}U_{\nu}$, respectively [see Eq.~(\ref{PM})].
From Eq.~(\ref{charged-lepton}) the hermitian square of the mass matrix for charged lepton $\mathcal{M}_{\ell}\mathcal{M}^{\dag}_{\ell}$ can be obtained. From Eqs.~(\ref{charged-lepton}) and (\ref{Vl}) the mixing angles and phases can be expressed in terms of Eq.~(\ref{ass}) as
 \begin{eqnarray}
 \theta^{\ell}_{3}\simeq\frac{1}{\sqrt{2}}\frac{Y^{\mu}_{2}}{Y_{\mu}}\frac{v_{\chi}}{\Lambda}~,\qquad\qquad\phi^{\ell}_{3}\simeq\frac{\varphi}{2}+\frac{1}{2}\arg\left(i-1\right)~,
 \label{lepMixing}
 \end{eqnarray}
and $\phi^{\ell}_{1}=\phi^{\ell}_{2}=0, \theta^{\ell}_{1}=\theta^{\ell}_{2}=0$, where the parameters $Y^{\mu}_{2}$, $Y_{\mu}$ are positive real numbers of order unity.
Then, from the above relation the mass squared eigenvalues are written, in a good approximation, as
 \begin{eqnarray}
 V^{\ell\dag}_{L}\mathcal{M}_{\ell}\mathcal{M}^{\dag}_{\ell}V^{d(\ell)}_{L} \equiv{\rm Diag}\left(m^{2}_{e},~m^{2}_{\mu},~m^{2}_{\tau}\right)
 \simeq{\rm Diag}\left(\frac{1}{2}Y^{\mu2}_{2}v^{2}_{H_2}\left(\frac{v_{\chi}}{\Lambda}\right)^{2},~\frac{1}{2}Y^{2}_{\mu}v^{2}_{H_2},~ \frac{1}{2}y^{2}_{\tau}v^{2}_{\eta}\right)~.
 \label{lepMass}
 \end{eqnarray}
Due to the ratio of the measured values $m_{e}/m_{\mu}\simeq\lambda^{3}$ and $m_{\mu}/m_{\tau}\simeq\lambda^{2}$, they can be expressed in terms of Eq.~(\ref{charged-lepton}) as $m_{e}/m_{\mu}\simeq Y^{\mu}_{2}v_{\chi}/(Y_{\mu}\Lambda)$ and $m_{\mu}/m_{\tau}\simeq Y_{\mu}v_{H_2}/(y_{\tau}v_{\eta})$. And,
one can express
 \begin{eqnarray}
 \frac{v_{\chi}}{\Lambda}\frac{Y^{\mu}_{2}}{\sqrt{2}Y_{\mu}}=A_{\ell}\lambda^{3}~,
 \label{lepMixing2}
 \end{eqnarray}
where the parameter $A_{\ell}$ is positive real number of order unity. Then, we can obtain the mixing matrix $V^{\ell}_{L}$ of the charged leptons: under the constraint of unitarity up to ${\cal O}(\lambda^{3})$, it can be written as
 \begin{eqnarray}
 V^{\ell}_{L}=
 {\left(\begin{array}{ccc}
 1 & A_{\ell}\lambda^{3} e^{i\phi^{\ell}_{3}} & 0  \\
 - A_{\ell}\lambda^{3} e^{-i\phi^{\ell}_{3}} &  1 & 0 \\
 0 & 0 & 1
 \end{array}\right)}P_{\ell}+{\cal O}(\lambda^{4})~.
 \label{CH}
 \end{eqnarray}
It indicates that the effect of mixing in the charged-lepton sector to the PMNS matrix is at least less than $\lambda^{3}\approx0.2^{\circ}$, and its contribution to the PMNS matrix is negligible because of the relatively large reactor angle $\theta_{13}$ measured in Daya Bay and RENO experiments~\cite{DayaRENO}.

\section{PMNS and CKM mixing matrices and Mass spectra}

\subsection{Quark Sector}
In the weak eigenstate basis, the mass terms in Eqs.~(\ref{lagrangianDown}) and (\ref{lagrangianUP}) and the charged gauge interactions can be written as
 \begin{eqnarray}
 -{\cal L}^{q}_{mW}
 &=& \overline{q^{u}_{L}}\mathcal{M}_{u}q^{u}_{R}+\overline{q^{d}_{L}}\mathcal{M}_{d}q^{d}_{R}
  +\frac{g}{\sqrt{2}}W^{+}_{\mu} ~\overline{q^{u}_{L}}\gamma^{\mu}q^{d}_{L} + {\rm h.c.} ~.
 \label{lagrangianA}
 \end{eqnarray}
Let us first consider the quark sector.
From Eq.~(\ref{lagrangianA}), to diagonalize the up- and down-type quark mass matrices such that
  \begin{align}
 V^{f\dag}_{L}\mathcal{M}_{f}V^{f}_{R} = {\rm Diag}(m_{f_1},m_{f_2},m_{f_3})~,
 \end{align}
we can rotate the quark fields from the weak eigenstates to the mass eigenstates:
\begin{eqnarray}
 q^{u(d)}_{L}\rightarrow V^{u(d)\dag}_{L}q^{u(d)}_{L}~, \qquad \quad q^{u(d)}_{R}\rightarrow V^{u(d)\dag}_{R}q^{u(d)}_{R}~.
 \label{rebasing}
\end{eqnarray}
Then, from the charged current terms in Eq.~(\ref{lagrangianA}), we obtain the CKM matrix
\begin{align}
V_{\rm CKM}=V^{u\dag}_{L}V^{d}_{L}~.
 \label{CKM}
\end{align}
From Eqs.~(\ref{DL}) and (\ref{UL}) we can obtain directly $V^{u\dag}_{L}V^{d}_{L}$, and
by recasting the result with the transformations $d\rightarrow de^{i\xi'}$, $c\rightarrow ce^{i(\phi^{d}_{3}-\xi+\xi')}$, $s\rightarrow se^{i(\phi^{d}_{3}-\xi)}$, $t\rightarrow te^{i(\phi^{d}_{3}+\frac{\varphi}{2}+\frac{\pi}{2}-\xi)}$ and $b\rightarrow be^{i(\phi^{d}_{3}+\frac{\varphi}{2}+\frac{\pi}{2}-\xi)}$, we can rewrite the CKM matrix as~\footnote{Note here that since the matrix in Eq.~(\ref{CKM1}) has the phase $\varphi$ dependence which is from $\chi$ field, when $C_{u}\rightarrow\lambda^{n}$ with $n\geq1$ ($n$: integer) one can re-parameterize and obtain the CKM matrix.}
 \begin{eqnarray}
 V_{\rm CKM}&=&
 {\left(\begin{array}{ccc}
 1 -\frac{\lambda^{2}}{2}+C_{u}\lambda^{3}\sin(\phi^{d}_{3}-\frac{\varphi}{2}) & \lambda-C_{u}\lambda^{2}\sin(\phi^{d}_{3}-\frac{\varphi}{2})  &  B \lambda^{3} e^{-i(\phi^{d}_{3}+\frac{\pi}{2}-\xi)} \\
 -\lambda+C_{u}\lambda^{2}\sin(\phi^{d}_{3}-\frac{\varphi}{2}) &  1-\frac{\lambda^{2}}{2}+C_{u}\lambda^{3}\sin(\phi^{d}_{3}-\frac{\varphi}{2}) & A \lambda^{2}e^{-i\xi'} \\
  (A-Be^{i(\phi^{d}_{3}+\frac{\pi}{2}-\xi)}) \lambda^{3}e^{-i\xi'}
  &  -A \lambda^{2} & 1
 \end{array}\right)}\nonumber\\
 &+&{\cal O}(\lambda^{4})~.
 \label{CKM1}
 \end{eqnarray}
where~\footnote{Taking into account the FCNCs in the up-type quarks we may approximate $A_{u},B_{u}, C_{u}\rightarrow0$ [see below Eq.~(\ref{UL})].} $A=A_{d}-A_{u}$, $B=B_{d}-B_{u}$, $\xi'\simeq C_{u}\lambda^{3}\cos(\frac{\varphi}{2}-\phi^{d}_{3})$ and $\xi\simeq C_{u}\lambda\cos(\frac{\varphi}{2}-\phi^{d}_{3})$.
If one set $C_{u}\rightarrow0$ which can be realized by $C_{u}\sim \frac{Y^{a}_{u}}{2Y_{u}}+\frac{Y^{s}_{u}}{3Y_{u}}\simeq\lambda^{n}$ with $n\geq1$, and
 \begin{eqnarray}
  Be^{-i(\phi^{d}_{3}+\frac{\pi}{2}-\xi)}=A(\rho+i\eta)~,
 \end{eqnarray}
then one can obtain the realistic CKM matrix in the Wolfenstein parametrization~\cite{Wolfenstein:1983yz} given by
 \begin{eqnarray}
 V_{\rm CKM}=
 {\left(\begin{array}{ccc}
 1 -\frac{\lambda^{2}}{2} & \lambda  &  A \lambda^{3} (\rho+i\eta) \\
 -\lambda &  1-\frac{\lambda^{2}}{2} & A \lambda^{2} \\
  A \lambda^{3}(1-\rho+i\eta)
  &  -A \lambda^{2} & 1
 \end{array}\right)}+{\cal O}(\lambda^{4})~.
 \label{CKM2}
 \end{eqnarray}
As reported in Ref.~\cite{ckmfitter} the best-fit values of the parameters $\lambda$, $A$, $\bar{\rho}$,
$\bar{\eta}$ with $1\sigma$ errors are
 \begin{eqnarray}
  \lambda &=& \sin\theta_{C}=0.22543\pm0.00077~,~~~~~A=0.812^{+0.013}_{-0.027}~, \nonumber\\
  \bar{\rho} &=& 0.144\pm0.025~,~~~~~~~~~~~~~~~~~~~~~~~\bar{\eta}=0.342^{+0.016}_{-0.015}~,
 \end{eqnarray}
where $\bar{\rho}=\rho(1-\lambda^{2}/2)$ and $\bar{\eta}=\eta(1-\lambda^{2}/2)$. The effects caused by
$CP$ violation are always proportional to the Jarlskog invariant~\cite{Jarlskog:1985ht}, defined as
$J^{\rm quark}_{CP} = {\rm Im}[V_{ud}V_{cs}V^{\ast}_{us}V^{\ast}_{cd}] \simeq A^{2} \lambda^{6}\eta$
whose value is $2.96^{+0.18}_{-0.17}\times10^{-5}$ at $1\sigma$ level~\cite{ckmfitter}.

From Eqs.~(\ref{downMass}) and (\ref{up-type mass}), the observed quark masses respect the following relation
 \begin{eqnarray}
 &&m_{d}:~m_{s}:~m_{b}
 \simeq v_{H_2}Y_{d}\sqrt{\frac{\lambda v_{\chi}}{\sqrt{2}\Lambda}\left\{Y_1\sin(\rho_{12}-\varphi)+ Y_{2}\cos(\rho_{12}-\varphi)\right\}}:~\frac{1}{2}Y_{d}v_{H_2}:~ \frac{1}{\sqrt{2}}y_{b}v_{\Psi}~,\nonumber\\
 &&m_{u}:~m_{c}:~ m_{t}\simeq \frac{1}{\sqrt{2}}Y_{u}v_{G_2}:~\frac{1}{\sqrt{2}}Y_{u}v_{G_1}:~\frac{1}{\sqrt{2}}y_{t}v_{\Phi}~.
 \end{eqnarray}

\subsection{Lepton Sector}
The mass terms in Eqs.~(\ref{lagrangiannu1}) and (\ref{lagrangianChLep}) and the charged gauge interactions in the weak eigenstate basis can be written in (block) matrix form as, using $\overline{N^{c}_R} m_D \nu^{c}_L = \overline{\nu_L} m_D^T N_R $,
 \begin{eqnarray}
 -{\cal L}^{\ell}_{mW} &=& \frac{1}{2}\overline{N^{c}_{R}}M_{R}N_{R}
 +\overline{\nu_{L}}m_{D}N_{R}+\overline{\ell_{L}}\mathcal{M}_{\ell}\ell_{R}+\frac{g}{\sqrt{2}}W^{-}_{\mu}\overline{\ell_{L}}\gamma^{\mu}\nu_{L}+\text{h.c.}~ \\
 &=& \frac{1}{2} \begin{pmatrix} \overline{\nu_L} & \overline{N^{c}_R} \end{pmatrix} \begin{pmatrix} 0 & m_D \\ m_D^T & M_R \end{pmatrix} \begin{pmatrix} \nu^{c}_L \\ N_R \end{pmatrix} + \overline{\ell_{L}}\mathcal{M}_{\ell}\ell_{R}+\frac{g}{\sqrt{2}}W^{-}_{\mu}\overline{\ell_{L}}\gamma^{\mu}\nu_{L}+\text{h.c.}
 \label{lagrangianAnu}
 \end{eqnarray}
Here $\ell=(e,\mu,\tau)$, $\nu=(\nu_e,\nu_\mu,\nu_\tau)$, $N_R=(N_{R1},N_{R2},N_{R3})$.

To find the neutrino masses and mixing matrix we are to diagonalize the $6\times6$ matrix
\begin{align}
 \begin{pmatrix} 0 & m_D \\ m^{T}_D & M_R \end{pmatrix} .
\end{align}
We start by diagonalizing $M_R$. For this purpose, we perform a basis rotation $\widehat{N}_{R} = U^{\dag}_{R}N_{R}$,
so that the right-handed Majorana mass matrix $M_{R}$ becomes a diagonal matrix $\widehat{M}_R$ with real and positive mass eigenvalues $M_{1}=a M$, $M_{2}=M$ and $M_{3}=b M$,
 \begin{eqnarray}
  \widehat{M}_{R}&=& U^{T}_{R}M_{R}U_{R}=M~U^{T}_{R}{\left(\begin{array}{ccc}
 1+\frac{2}{3}\kappa e^{i\varphi} &  -\frac{1}{3}\kappa e^{i\varphi} &  -\frac{1}{3}\kappa e^{i\varphi} \\
 -\frac{1}{3}\kappa e^{i\varphi} &  \frac{2}{3}\kappa e^{i\varphi} &  1-\frac{1}{3}\kappa e^{i\varphi} \\
 -\frac{1}{3}\kappa e^{i\varphi} &  1-\frac{1}{3}\kappa e^{i\varphi} &  \frac{2}{3}\kappa e^{i\varphi}
 \end{array}\right)}U_{R}
= \begin{pmatrix} aM & 0 & 0 \\ 0 & M & 0 \\ 0 & 0 & bM \end{pmatrix},
  \label{heavy}
 \end{eqnarray}
where $\kappa=y_R^\nu v_{\chi}/M$. We find $a=\sqrt{1+\kappa^{2}+2\kappa\cos\varphi}$, $b=\sqrt{1+\kappa^{2}-2\kappa\cos\varphi}$, and a diagonalizing matrix
\begin{eqnarray}
  U_{R} = {\left(\begin{array}{ccc}
  \sqrt{\frac{2}{3}}  &  \frac{1}{\sqrt{3}}  &  0 \\
  -\frac{1}{\sqrt{6}}  &  \frac{1}{\sqrt{3}}  &  -\frac{1}{\sqrt{2}} \\
  -\frac{1}{\sqrt{6}} &  \frac{1}{\sqrt{3}}  &  \frac{1}{\sqrt{2}}
  \end{array}\right)}{\left(\begin{array}{ccc}
  e^{i\frac{\psi_1}{2}}  &  0  &  0 \\
  0  &  1  &  0 \\
  0  &  0  &  e^{i\frac{\psi_2}{2}}
  \end{array}\right)}~,
  \label{URN}
\end{eqnarray}
with phases
\begin{eqnarray}
 \psi_1 = \tan^{-1} \Big( \frac{-\kappa\sin\varphi}{1+\kappa\cos\varphi} \Big)
 ~~~{\rm and}~~~ \psi_2 = \tan^{-1} \Big( \frac{\kappa\sin\varphi}{1-\kappa\cos\varphi} \Big)~.
\label{alphs_beta}
\end{eqnarray}
As the magnitude of $\kappa$ defined in Eq.~(\ref{heavy}) decreases, the phases $\psi_{1,2}$ go to $0$ or $\pi$.
With the basis rotation $N_{R}\rightarrow U^{\dag}_{R}N_{R}$, the Dirac neutrino mass matrix gets modified to
 \begin{eqnarray}
 m_{D}\rightarrow\widetilde{m}_{D}=m_{D}U_{R}= \frac{v_{\Phi}e^{i\gamma}}{\sqrt{2}}y^{\nu}_{3}{\left(\begin{array}{ccc}
 1 & 0 & 0 \\
 0 & 0 & y_{2} \\
 0 & y_{1} & 0
 \end{array}\right)}U_{R}~,
 \label{mDT}
 \end{eqnarray}
where $y_{1}=y^{\nu}_{1}/y^{\nu}_{3}, y_{2}=y^{\nu}_{2}/y^{\nu}_{3}$.
At this point,
\begin{align}
-{\cal L}_{mW} &= \frac{1}{2} \begin{pmatrix} \overline{\nu_L} & \overline{\widehat{N}^{c}_R} \end{pmatrix} \begin{pmatrix} 0 & \widetilde{m}_D \\ \widetilde{m}_D^T & \widehat{M}_R \end{pmatrix} \begin{pmatrix} \nu^{c}_L \\ \widehat{N}_R \end{pmatrix} + \overline{\ell_{L}}\mathcal{M}_{\ell}\ell_{R}+\frac{g}{\sqrt{2}}W^{-}_{\mu}\overline{\ell_{L}}\gamma^{\mu}\nu_{L}+\text{h.c.}~.
 \label{lagrangianB}
\end{align}
Now we take the limit of large $M$ (seesaw mechanism) and focus on the mass matrix of the light neutrinos $M_{\nu}$,
\begin{align}
-{\cal L}_{mW} &= \frac{1}{2} \overline{\nu_L} \mathcal{M}_{\nu} \nu^{c}_L  + \overline{\ell_{L}}\mathcal{M}_{\ell}\ell_{R}+\frac{g}{\sqrt{2}}W^{-}_{\mu}\overline{\ell_{L}}\gamma^{\mu}\nu_{L}+\text{h.c.}+\text{terms in $N_R$}
\end{align}
with
 \begin{align}
 \mathcal{M}_{\nu} = - \widetilde{m}_D \, \widehat{M}_R^{-1} \, \widetilde{m}^T_D.
 \end{align}
We perform basis rotations from weak  to mass eigenstates in the leptonic sector,
\begin{eqnarray}
 \widehat{\ell}_{L} = V^{\ell\dag}_{L}\ell_{L}~,\qquad \widehat{\ell}_{R}= V^{\ell\dag}_{R}\ell_{R}~,\qquad \widehat{\nu}_{L} = U^{\dag}_{\nu}\nu_{L}~,
 \label{rebasing}
\end{eqnarray}
where $U_\nu, V_{L(R)}$ are unitary matrices chosen so as the matrices
\begin{align}
\widehat{\mathcal{M}}_{\nu} &= U^{\dag}_{\nu}  \mathcal{M}_\nu  U^*_{\nu} = - U_{\nu}^{\dag} m_D U_R   \widehat{M}_R^{-1} (U_{\nu}^{\dag} m_D U_R)^T~,\nonumber\\
\widehat{\mathcal{M}}_{\ell} &= V^{\ell\dag}_{L}  \mathcal{M}_\ell  V_{R}
\end{align}
are diagonal. From Eqs.~(\ref{lepMass}) and (\ref{lepMixing2}) the observed charged lepton masses respect
 \begin{eqnarray}
 m_{e}:~m_{\mu}:~m_{\tau}
 \simeq A_{\ell}\lambda^{3}Y_{\mu}v_{H_2}:~\frac{Y_{\mu}}{\sqrt{2}}v_{H_2}:~ \frac{y_{\tau}}{\sqrt{2}}v_{\eta}~.
 \end{eqnarray}
 And from the charged current term in Eq.~(\ref{lagrangianB}) we obtain the lepton mixing matrix $U_{\rm PMNS}$ as
\begin{align}
U_{\rm PMNS}=V^{\ell\dag}_{L}U_{\nu}.
 \label{PM}
\end{align}
The matrix $U_{\rm PMNS}$ can be written in terms of three mixing angles and three $CP$-odd phases (one for the Dirac neutrinos and two for the Majorana neutrinos) as \cite{PDG}
\begin{eqnarray}
  U_{\rm PMNS}
  &=&{\left(\begin{array}{ccc}
   c_{13}c_{12} & c_{13}s_{12} & s_{13}e^{-i\delta_{CP}} \\
   -c_{23}s_{12}-s_{23}c_{12}s_{13}e^{i\delta_{CP}} & c_{23}c_{12}-s_{23}s_{12}s_{13}e^{i\delta_{CP}} & s_{23}c_{13}  \\
   s_{23}s_{12}-c_{23}c_{12}s_{13}e^{i\delta_{CP}} & -s_{23}c_{12}-c_{23}s_{12}s_{13}e^{i\delta_{CP}} & c_{23}c_{13}
   \end{array}\right)}P_{\nu}~,
 \label{rebasing1}
\end{eqnarray}
where $P_{\nu}={\rm Diag}(e^{-i\varphi_{1}/2},e^{-i\varphi_{2}/2},1)$, and $s_{ij}\equiv \sin\theta_{ij}$ and $c_{ij}\equiv \cos\theta_{ij}$.

After seesawing, the light neutrino mass matrix is given by~\footnote{The neutrino mass matrix form given by Eq.~(\ref{meff}) is different from the one given in Ref.~\cite{Ahn:2012cg} due to the opposite sign in front of $\frac{3e^{i\psi_{2}}}{2b}$ in 2-3 sector of the light neutrino mass matrix. So, it makes a difference in numerical results.}
 \begin{eqnarray}
  \mathcal{M}_{\nu} &=& -\widetilde{m}_{D}\widehat{M}^{-1}_{R}\widetilde{m}^{T}_{D} \nonumber\\
  &=&e^{2i\gamma} m_{0}
   {\left(\begin{array}{ccc}
   1+\frac{2e^{i\psi_{1}}}{a} & (1-\frac{e^{i\psi_{1}}}{a})y_{2} & (1-\frac{e^{i\psi_{1}}}{a})y_{1} \\
   (1-\frac{e^{i\psi_{1}}}{a})y_{2} & (1+\frac{e^{i\psi_{1}}}{2a}+\frac{3e^{i\psi_{2}}}{2b})y^{2}_{2} & (1+\frac{e^{i\psi_{1}}}{2a}-\frac{3e^{i\psi_{2}}}{2b})y_{1}y_{2}  \\
   (1-\frac{e^{i\psi_{1}}}{a})y_{1} & (1+\frac{e^{i\psi_{1}}}{2a}-\frac{3e^{i\psi_{2}}}{2b})y_{1}y_{2} & (1+\frac{e^{i\psi_{1}}}{2a}+\frac{3e^{i\psi_{2}}}{2b})y^{2}_{1}
   \end{array}\right)}~,
  \label{meff}
 \end{eqnarray}
where we have defined an overall scale $m_{0}=v^{2}_{\Phi}y^{\nu2}_{3}/(6M)$ for the light neutrino masses. And the overall phase can be rotated away by redefining the light neutrino field.
The mass matrix $\mathcal{M}_{\nu}$ is diagonalized by the mixing matrix $U_{\nu}$,
 \begin{eqnarray}
  \mathcal{M}_{\nu} &=& U_{\nu} ~{\rm Diag}(m_{\nu_1},m_{\nu_2},m_{\nu_3})~ U^{T}_{\nu} .
 \end{eqnarray}
Here $m_{\nu_i}$ $(i = 1,2,3)$ are the light neutrino masses. Interestingly, the mixing matrix $U_{R}$ in Eq.~(\ref{URN}) reflects an exact TBM.
Therefore Eq.~(\ref{meff}) directly indicates that there could be deviations from the exact TBM if the Dirac neutrino Yukawa couplings do not have the same magnitude.
In the limit $y^\nu_1=y^\nu_2$, the mass matrix in Eq.~(\ref{meff}) acquires a $\mu$--$\tau$ symmetry~\cite{mutau} that leads to $\theta^{\nu}_{13}=0$ and $\theta^{\nu}_{23}=-\pi/4$. Moreover, in the limit  $y^\nu_1=y^\nu_2=y^\nu_3$ ($y_{1}, y_{2}\rightarrow1$), the  mass matrix~(\ref{meff}) gives the TBM angles and the corresponding mass eigenvalues:
 \begin{eqnarray}
 \theta^{\nu}_{1}&=& -\frac{\pi}{4}~,~\qquad \theta^{\nu}_{2}=0~,\qquad ~~\theta^{\nu}_{3}= \sin^{-1}\left(\frac{1}{\sqrt{3}}\right)~,\nonumber\\
 m_{\nu_1}&=& \frac{3m_{0}}{a}~,\qquad m_{\nu_2}=3m_{0}~,\qquad m_{\nu_3}= \frac{3m_{0}}{b}~.
 \label{TBM1}
 \end{eqnarray}
These mass eigenvalues are disconnected from the mixing angles. The neutrino texture in Eq.~(\ref{meff}) provides naturally the mildness of neutrino masses, because the components giving the TBM are multiplied and constrained by neutrino Yukawa couplings.

Due to in general $y_{1},y_{2}\neq1$, there are deviations from their TBM values.
Moreover, recent neutrino data, {\it i.e.} $\theta_{13}\neq0$, require deviations of $y_{1,2}$ from unity because the contribution of $V^{\ell}_{L}$ in Eq.~(\ref{CH}) from the charged-lepton sector is expected to be small, leading to a possibility to search for $CP$ violation in neutrino oscillation experiments.
To diagonalize the above mass matrix Eq.~(\ref{meff}), we consider the hermitian matrix $\mathcal{M}_{\nu}\mathcal{M}^{\dag}_{\nu}$, from which we obtain the masses and mixing angles:
 \begin{eqnarray}
 {\cal M}_{\nu}{\cal M}^{\dag}_{\nu}=\left(\begin{array}{ccc}
  A & B & C \\
  B^{\ast} & F & |G|e^{i\phi^{\nu}_{1}} \\
  C^{\ast} & |G|e^{-i\phi^{\nu}_{1}} & K
  \end{array}\right)=U_{\nu}~{\rm Diag}(m^{2}_{\nu_1},m^{2}_{\nu_2},m^{2}_{\nu_3})~U^{\dag}_{\nu}~,
 \label{MM}
 \end{eqnarray}
where
 \begin{eqnarray}
 A&=&p+q+2g_{1}~,\quad F=y^{2}_{2}\left(p+\frac{q}{4}+r-g_1-g_2\right)~,\quad K=y^{2}_{1}\left(p+\frac{q}{4}+r-g_1+g_2\right)~,\nonumber\\
 B&=&y_{2}\left(p-\frac{q}{2}+\frac{g_1}{2}-g_3+i\frac{3(g_4+g_5)}{2a}\right)~,~ C=y_{1}\left(p-\frac{q}{2}+\frac{g_1}{2}+g_3+i\frac{3(g_4-g_5)}{2a}\right)~,\nonumber\\
 G&=&y_{1}y_{2}\left(p+\frac{q}{4}-r-g_1+ig_6\right)~,
 \label{MM1}
 \end{eqnarray}
with
 \begin{eqnarray}
 p&=&m^{2}_{0}(1+y^{2}_{1}+y^{2}_{2})~,\qquad\quad q=m^{2}_{0}\frac{4+y^{2}_{1}+y^{2}_{2}}{a^{2}}~,\qquad\quad r=9m^{2}_{0}\frac{y^{2}_{1}+y^{2}_{2}}{4b^{2}}~,\nonumber\\
 g_1&=&m^{2}_{0}\cos\psi_1\frac{2-y^{2}_{1}-y^{2}_{2}}{a}~,\qquad\qquad\qquad g_2=3m^{2}_{0}(y^{2}_{1}-y^{2}_{2})\frac{\cos\gamma+2a\cos\psi_2}{2ab}~,\nonumber\\
 g_3&=&3m^{2}_{0}(y^{2}_{1}-y^{2}_{2})\frac{a\cos\psi_2-\cos\gamma}{2ab}~,\qquad~ g_{4}=m^{2}_{0}(2-y^{2}_{1}-y^{2}_{2})\sin\psi_1~,\nonumber\\
 g_5&=&m^{2}_{0}(y^{2}_{1}-y^{2}_{2})\frac{a\sin\psi_2+\sin\gamma}{b}~,\qquad\quad g_6=3m^{2}_{0}(y^{2}_{1}-y^{2}_{2})\frac{\sin\gamma-2a\sin\psi_2+}{2ab}~,
 \label{MM2}
 \end{eqnarray}
and $\gamma\equiv\psi_1-\psi_2=\tan^{-1}\left(\frac{2\kappa\sin\varphi}{\kappa^{2}-1}\right)$.
In the limit of $y_{1},y_{2}\rightarrow1$ the parameters in Eq.~(\ref{MM2}) behave as $p\rightarrow3m^{2}_{0},q\rightarrow6m^{2}_{0}/a^{2},r\rightarrow2m^{2}_{0}/b^{2}$ and $g_{i}\rightarrow0$. So, as expected, the matrix in Eq.~(\ref{MM}) gives the TBM values.
Similarly Eq.~(\ref{Vl}), we have three mixing angles ($\theta^{\nu}_{1}, \theta^{\nu}_{2},\theta^{\nu}_{3}$), three phases ($\phi^{\nu}_{1},\phi^{\nu}_{2},\phi^{\nu}_{3}$), and the three mass-squared eigenvalues. In turn, this mixing matrix $U_{\nu}$ becomes one of the mixing matrix composing the PMNS matrix.
To see how the neutrino mass matrix given by Eq.(\ref{meff}) can lead to deviations from their TBM values,
we first introduce three small quantities $\varepsilon_{i},~(i=1,2,3)$, which are responsible for the deviations of the $\theta_{j}$ from their TBM values:
 \begin{eqnarray}
  \theta^{\nu}_{1}=-\frac{\pi}{4}+\varepsilon_{1}~, \qquad\theta^{\nu}_{2}=\varepsilon_{2}~, \qquad\theta^{\nu}_{3}= \sin^{-1}\left(\frac{1}{\sqrt{3}}\right)+\varepsilon_{3}~.
 \end{eqnarray}
Then the mixing matrix $U_{\nu}$ up to order $\varepsilon_{i}$ can be written as
 \begin{eqnarray}
 U_{\nu}&=&{\left(\begin{array}{ccc}
 \frac{\sqrt{2}-\varepsilon_{3}}{\sqrt{3}} &  \frac{1+\varepsilon_{3}\sqrt{2}}{\sqrt{3}}e^{i\phi^{\nu}_{3}} &  \varepsilon_{2}e^{i\phi^{\nu}_{2}} \\
 -\frac{(1+\varepsilon_{1}+\varepsilon_{3}\sqrt{2})e^{-i\phi^{\nu}_{3}}}{\sqrt{6}}+\frac{\varepsilon_{2}e^{i(\phi^{\nu}_{1}-\phi^{\nu}_{2})}}{\sqrt{3}} &  \frac{\sqrt{2}+\varepsilon_{1}\sqrt{2}-\varepsilon_{3}}{\sqrt{6}}+\frac{\varepsilon_{2}e^{i(\phi^{\nu}_{1}-\phi^{\nu}_{2}+\phi^{\nu}_{3})} }{\sqrt{6}} &  \frac{(-1+\varepsilon_{1})e^{i\phi^{\nu}_{1}}}{\sqrt{2}} \\
 -\frac{(1-\varepsilon_{1}-\varepsilon_{3}\sqrt{2})e^{-i(\phi^{\nu}_{1}+\phi^{\nu}_{3})}}{\sqrt{6}}-\frac{\varepsilon_{2}e^{-i\phi^{\nu}_{2}}}{\sqrt{3}} &  \frac{(\sqrt{2}-\varepsilon_{3}-\sqrt{2}\varepsilon_{1})e^{-i\phi^{\nu}_{1}}}{\sqrt{6}}-\frac{\varepsilon_{2}e^{i(\phi^{\nu}_{3}-\phi^{\nu}_{2})}}{\sqrt{6}} &  \frac{1+\varepsilon_{1}}{\sqrt{2}}
 \end{array}\right)}P_{\nu}\nonumber\\
 &+&{\cal O}(\varepsilon^{2}_{i})~.
 \label{Unu}
 \end{eqnarray}
 
Now, the straightforward calculation with the general
parametrization of $U_{\nu}$ in Eq.~(\ref{Vl}) leads to the expressions for the masses and mixing parameters
 \begin{eqnarray}
  \tan\theta^{\nu}_{1}&=&\frac{{\rm Im}[C]\sin\phi^{\nu}_{2}-{\rm Re}[C]\cos\phi^{\nu}_{2}}{{\rm Im}[B]\cos(\phi^{\nu}_{1}-\phi^{\nu}_{2})+{\rm Re}[B]\sin(\phi^{\nu}_{1}-\phi^{\nu}_{2})}~, \qquad \phi^{\nu}_{1}=\arg(G)~,\nonumber\\ \tan2\theta^{\nu}_{2}&=&2\frac{|c^{\nu}_{1}C+e^{i\phi^{\nu}_{1}}s^{\nu}_{1}B|}{\lambda_{3}-A}~,\qquad\phi^{\nu}_{2}=\arg\left(c^{\nu}_{1}C+e^{i\phi^{\nu}_{1}}s^{\nu}_{1}B\right)~,\nonumber\\
  \tan2\theta^{\nu}_{3}&=&2\frac{|Z|}{\lambda_{2}-\lambda_{1}}~,~\quad \phi^{\nu}_{3}=\arg(Z)~,
 \label{Theta1312}
 \end{eqnarray}
where $c^{\nu}_{i}=\cos\theta^{\nu}_{i}, s^{\nu}_{i}=\sin\theta^{\nu}_{i}$, and
 \begin{eqnarray}
  \lambda_{1}&=&Ac^{\nu2}_{2}-|c^{\nu}_{1}C+e^{i\phi^{\nu}_{1}}s^{\nu}_{1}B|\sin2\theta^{\nu}_{2}+\lambda_{3}s^{\nu2}_{2}~,\nonumber\\
  \lambda_{2}&=&Fc^{\nu2}_{1}-|G|\sin2\theta^{\nu}_{1}+Ks^{\nu2}_{1}~,\qquad\qquad
  \lambda_{3}=Kc^{\nu2}_{1}+|G|\sin2\theta^{\nu}_{1}+Fs^{\nu2}_{1}~,\nonumber\\
  Z&=& c^{\nu}_{2}(c^{\nu}_{1}B-e^{-i\phi^{\nu}_{1}}s^{\nu}_{1}C)+s^{\nu}_{2}e^{i(\phi^{\nu}_{2}-\phi^{\nu}_{1})}\left(\sin2\theta^{\nu}_{1}\frac{K-F}{2}-|G|\cos2\theta^{\nu}_{1}\right)~.
 \label{para1}
 \end{eqnarray}
And the squared-mass eigenvalues are given by
 \begin{eqnarray}
    m^{2}_{\nu_1} = \frac{\lambda_{1}c^{\nu2}_{3}-\lambda_{2}s^{\nu2}_{3}}{\cos2\theta^{\nu}_{3}}~,\quad
    m^{2}_{\nu_2} = \frac{\lambda_{2}c^{\nu2}_{3}-\lambda_{1}s^{\nu2}_{3}}{\cos2\theta^{\nu}_{3}}~,\quad
    m^{2}_{\nu_3}=\lambda_{3}+|c^{\nu}_{1}C+e^{i\phi^{\nu}_{1}}s^{\nu}_{1}B|\tan\theta^{\nu}_{2}~.
 \label{eigenvalueNu}
 \end{eqnarray}
As is well-known, because of the observed hierarchy $\Delta m^{2}_{\rm Atm}\gg\Delta m^{2}_{\rm Sol}\equiv m^{2}_{2}-m^{2}_{1}>0$, and the requirement of a Mikheyev-Smirnov-Wolfenstein resonance for solar neutrinos, there are two possible neutrino mass spectra: (i) the normal mass hierarchy (NMH) $m_{1}<m_{2}<m_{3}$, and (ii) the inverted mass hierarchy (IMH) $m_{3}<m_{1}<m_{2}$.
The solar and atmospheric mass-squared differences are given by
\begin{eqnarray}
 \Delta m^{2}_{\rm sol}\equiv m^{2}_{\nu_2}-m^{2}_{\nu_1}
  &=& \frac{2\left|Z\right|}{\sin2\theta^{\nu}_{3}} ~,\qquad
 \Delta m^{2}_{\rm atm}\equiv\left\{
                               \begin{array}{ll}
                                 m^{2}_{\nu_3}-m^{2}_{\nu_1}, & \hbox{for NMH} \\
                                 m^{2}_{\nu_2}-m^{2}_{\nu_3}, & \hbox{for IMH}
                               \end{array}~,
                             \right.
 \label{deltam2}
\end{eqnarray}
which are constrained by the neutrino oscillation experimental results. We will discuss it numerically in the next section.

Plugging Eqs.~(\ref{CH}) and (\ref{Unu}) into Eq.~(\ref{PM}), the PMNS matrix is recast to
 \begin{eqnarray}
  U_{\rm PMNS}=
 {\left(\begin{array}{ccc}
 U_{\nu11}-A_{\ell}\lambda^{3}e^{i\phi^{\ell}_{3}}U_{\nu21} & U_{\nu12}-A_{\ell}\lambda^{3}e^{i\phi^{\ell}_{3}}U_{\nu22} & U_{\nu13}-A_{\ell}\lambda^{3}e^{i\phi^{\ell}_{3}}U_{\nu23} \\
 U_{\nu21}+A_{\ell}\lambda^{3}e^{-i\phi^{\ell}_{3}}U_{\nu11} & U_{\nu22}+A_{\ell}\lambda^{3}e^{-i\phi^{\ell}_{3}}U_{\nu12} & U_{\nu23}+A_{\ell}\lambda^{3}e^{-i\phi^{\ell}_{3}}U_{\nu13} \\
 U_{\nu31} & U_{\nu32} & U_{\nu33}
 \end{array}\right) P_{\nu}}~,
 \label{PMNS2}
 \end{eqnarray}
where the phase $\phi^{\ell}_{3}$ is given as Eq.~(\ref{lepMixing}).
From Eq.~(\ref{PMNS2}), the neutrino mixing parameters can be displayed in terms of the standard parametrization~\cite{PDG} as
  \begin{eqnarray}
  \sin^{2}\theta_{12}&=&\frac{|U_{e2}|^{2}}{1-|U_{e3}|^{2}}~,\qquad
   \sin^{2}\theta_{23}=\frac{|U_{\mu3}|^{2}}{1-|U_{e3}|^{2}}~,\qquad
  \sin\theta_{13}=|U_{e3}|~.
 \label{mixing1}
 \end{eqnarray}
Leptonic $CP$ violation at low energies can be detected through the neutrino oscillations which are
sensitive to the Dirac $CP$-phase, but insensitive to the Majorana $CP$-phases in
$U_{\rm PMNS}$~\cite{Branco:2002xf}:
the Jarlskog invariant~\cite{Jarlskog:1985ht} is defined as
  \begin{eqnarray}
  J_{CP}\equiv-{\rm Im}[U^{\ast}_{e1}U_{e3}U_{\tau1}U^{\ast}_{\tau3}]=\frac{1}{8}\sin2\theta_{12}\sin2\theta_{13}\sin2\theta_{23}\cos\theta_{13}\sin\delta_{CP}~,
 \label{JCP}
 \end{eqnarray}
where $U_{\alpha j}$ is an element of the PMNS matrix in Eq.~(\ref{PMNS2}), with $\alpha=e,\mu,\tau$
corresponding to the lepton flavors and $j=1,2,3$ corresponding to the light neutrino mass eigenstates.
And by manipulation of Eqs.~(\ref{rebasing1}) and (\ref{JCP}) one can easily obtain the Dirac $CP$ phase :
 \begin{eqnarray}
  \delta_{CP}=-\arg\left(\frac{\frac{U^{\ast}_{e1}U_{e3}U_{\tau1}U^{\ast}_{\tau3}}{c_{12}c^{2}_{13}c_{23}s_{13}}+c_{12}c_{23}s_{13}}{s_{12}s_{23}}\right)~.
 \label{DeltaCP1}
 \end{eqnarray}
As expected, since the contributions of $V^{\ell}_{L}$ in Eq.~(\ref{CH}) to the PMNS matrix are negligible, {\it i.e.}, its effects ${\cal O}(\lambda^{3})$, we will consider $U_{\rm PMNS}\simeq U_{\nu}$ in numerical analysis.
\begin{figure}[t]
\begin{minipage}[t]{6.0cm}
\epsfig{figure=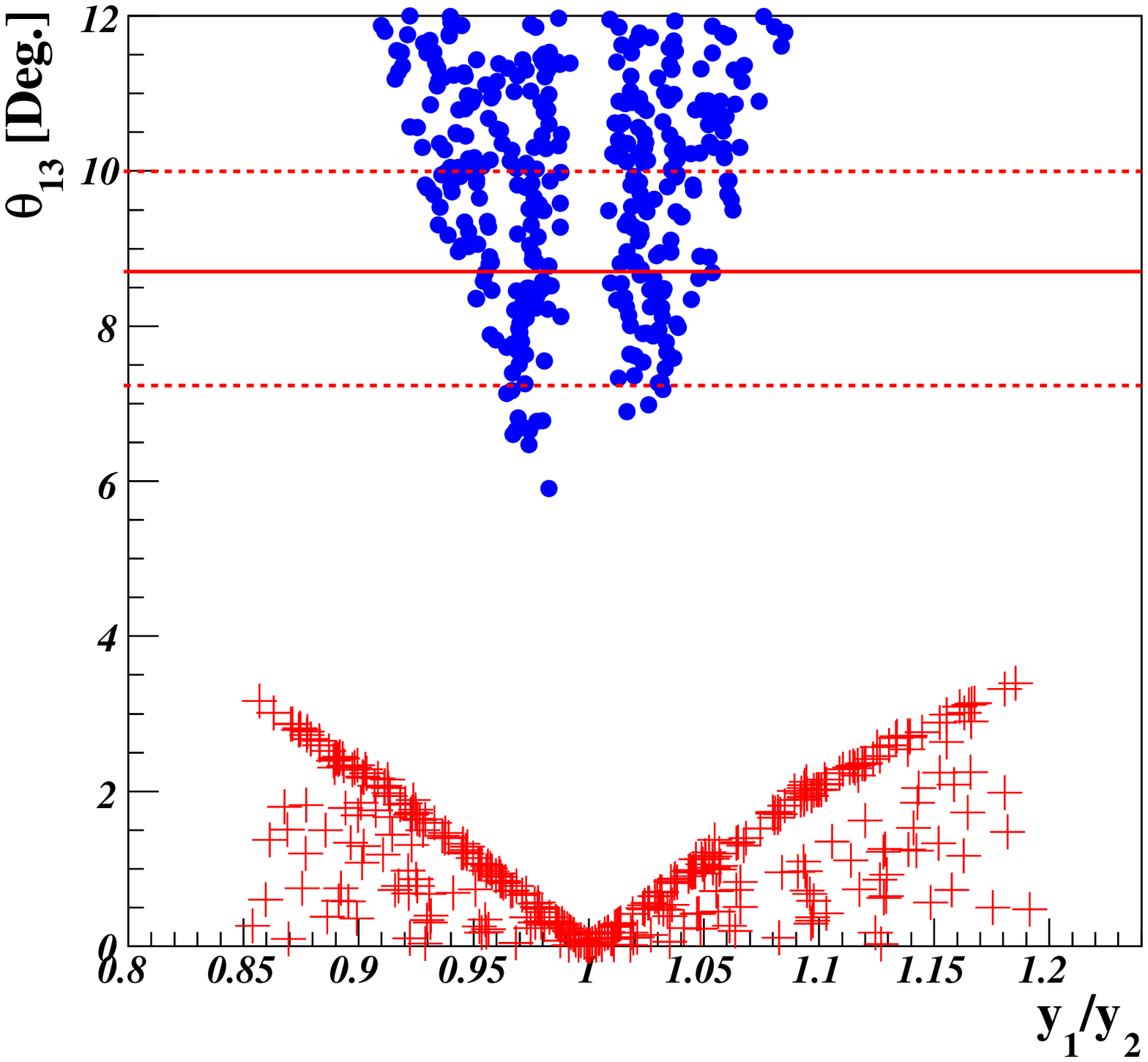,width=6.5cm,angle=0}
\end{minipage}
\hspace*{1.0cm}
\begin{minipage}[t]{6.0cm}
\epsfig{figure=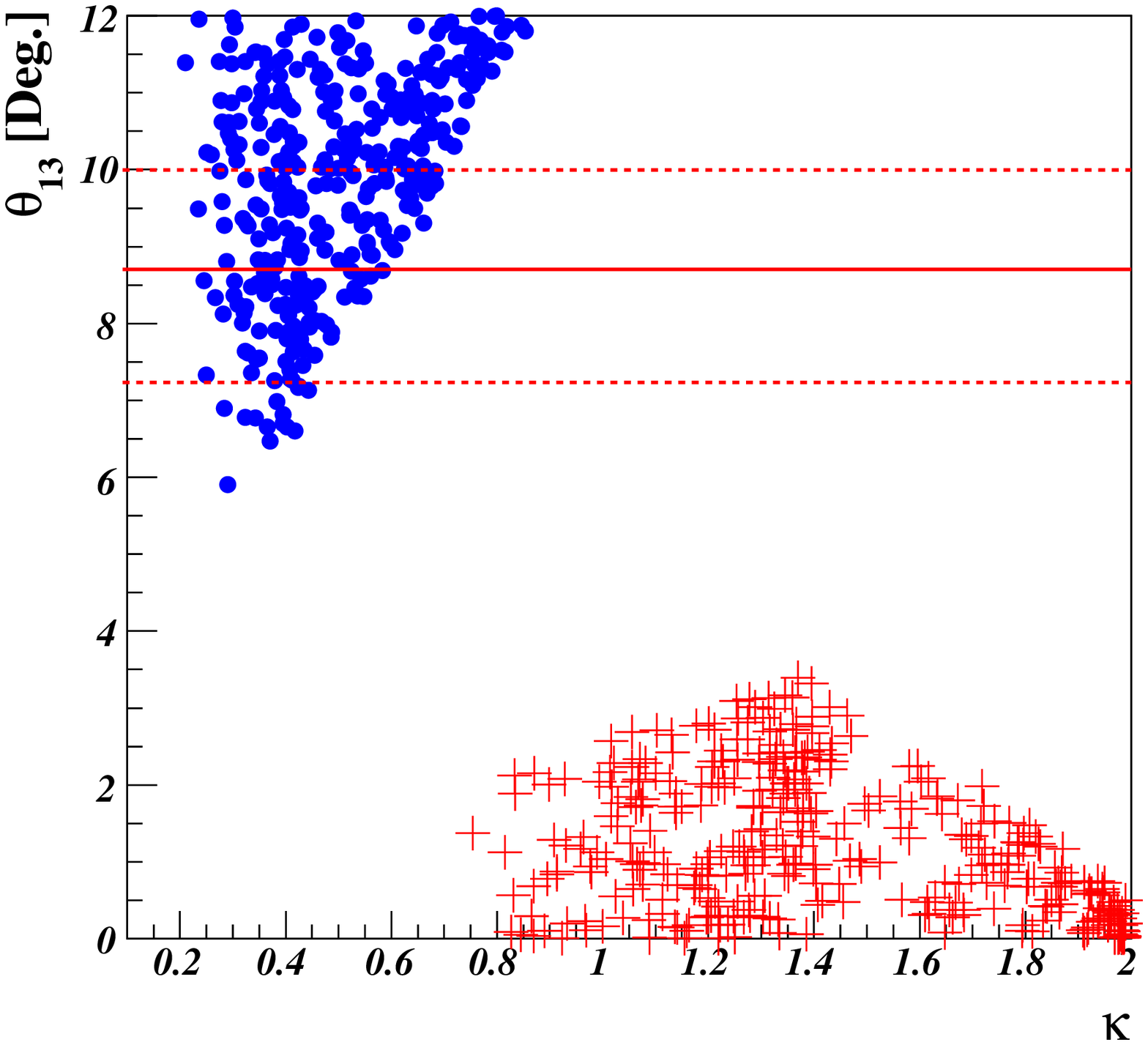,width=6.5cm,angle=0}
\end{minipage}
\caption{\label{FigA1}
The reactor mixing angle $\theta_{13}$ versus the ratio of third-to-second generation neutrino Yukawa couplings $y^{\nu}_{1}/y^{\nu}_{2}$ (left plot) and the parameter $\kappa=y^{\nu}_{R} v_\chi/M$ (right plot). The (red) crosses and (blue) dots represent the results for the inverted and the normal mass hierarchy, respectively.  The horizontal dotted (solid) lines in both plots indicate the upper and lower bounds on $\theta_{13}$ given in Eq.~(\ref{expnu}) at the $3\sigma$ level (best-fit value).}
\end{figure}

\section{Numerical Analysis}
Now we perform a numerical analysis for neutrinos using the linear algebra tools in Ref.~\cite{Antusch:2005gp}.

The mass matrices $\widetilde{m}_{D}$ and $\widehat{M}_{R}$ in Eq.~(\ref{meff}) contains seven parameters: $y^{\nu}_{3},v_{\Phi}, M, y_{1}, y_{2}, \kappa, \varphi$. The first three ($y^{\nu}_{3}$, $M$ and $v_{\Phi}$) lead to the overall  neutrino scale parameter $m_{0}$. The next four ($y_1,y_2,\kappa,\varphi$) give rise to the deviations from TBM as well as the $CP$ phases and corrections to the masse eigenvalues [see Eq.~(\ref{TBM1})].
Since we have a relation $v_{\chi}/\Lambda\sim\lambda^{2}$ in the charged fermion sector, for the cutoff scale $\Lambda=10^{15}$ GeV we take $M=10^{13}$ GeV and $v_{\Phi}=172\sqrt{2}$ GeV, for simplicity, as inputs. Since the neutrino masses are sensitive to the combination $m_{0}=v^{2}_{\Phi}|y^{\nu2}_{3}|/(6M)$, other choices of $M$ and $v_\Phi$ give identical results. Then the parameters $m_{0},y_{1},y_{2},\kappa,\varphi$ can be determined from the experimental results of three mixing angles, $\theta_{12},\theta_{13},\theta_{23}$, and the
two mass squared differences, $\Delta m^{2}_{\rm Sol}, \Delta m^{2}_{\rm Atm}$. In addition, the $CP$ phases $\delta_{CP},\varphi_{1,2}$ can be predicted after
determining the model parameters. (Here, we will not discuss the Majorana $CP$ phases $\varphi_{1,2}$.)
\begin{figure}[t]
\begin{minipage}[t]{6.0cm}
\epsfig{figure=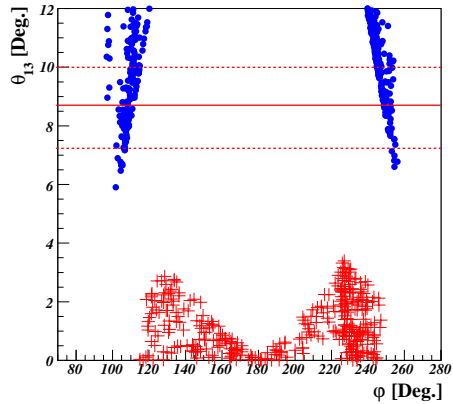,width=6.5cm,angle=0}
\end{minipage}
\caption{\label{FigA1p}
The reactor mixing angle $\theta_{13}$ versus the phase $\varphi$. The (red) crosses and (blue) dots represent the results for the inverted and normal mass hierarchy, respectively. The vertical dotted (solid) lines indicate the upper and lower bounds on $\theta_{13}$ given in Eq.~(\ref{expnu}) at the $3\sigma$ level (best-fit value).}
\end{figure}

\begin{figure}[t]
\begin{minipage}[t]{6.0cm}
\epsfig{figure=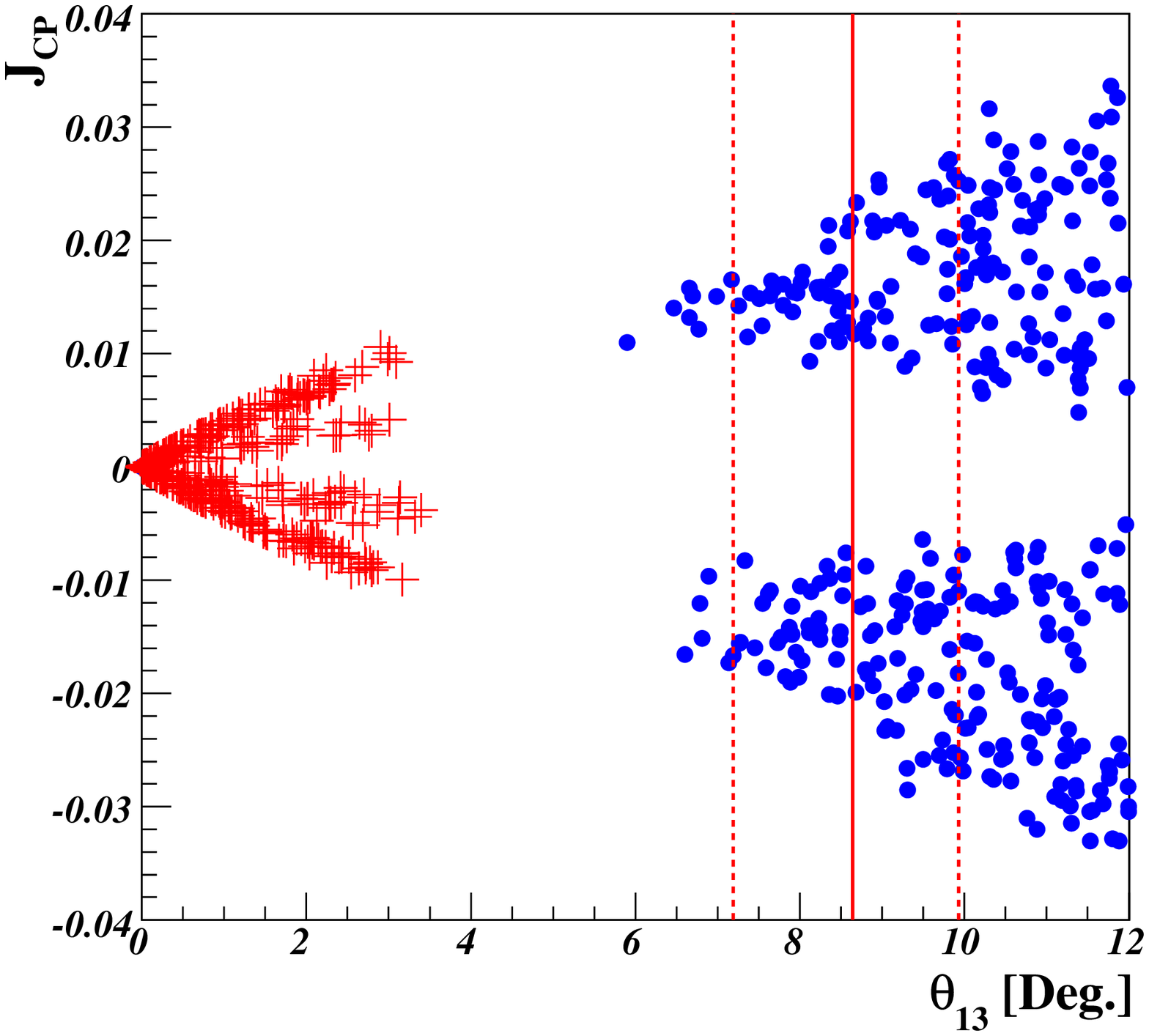,width=6.5cm,angle=0}
\end{minipage}
\hspace*{1.0cm}
\begin{minipage}[t]{6.0cm}
\epsfig{figure=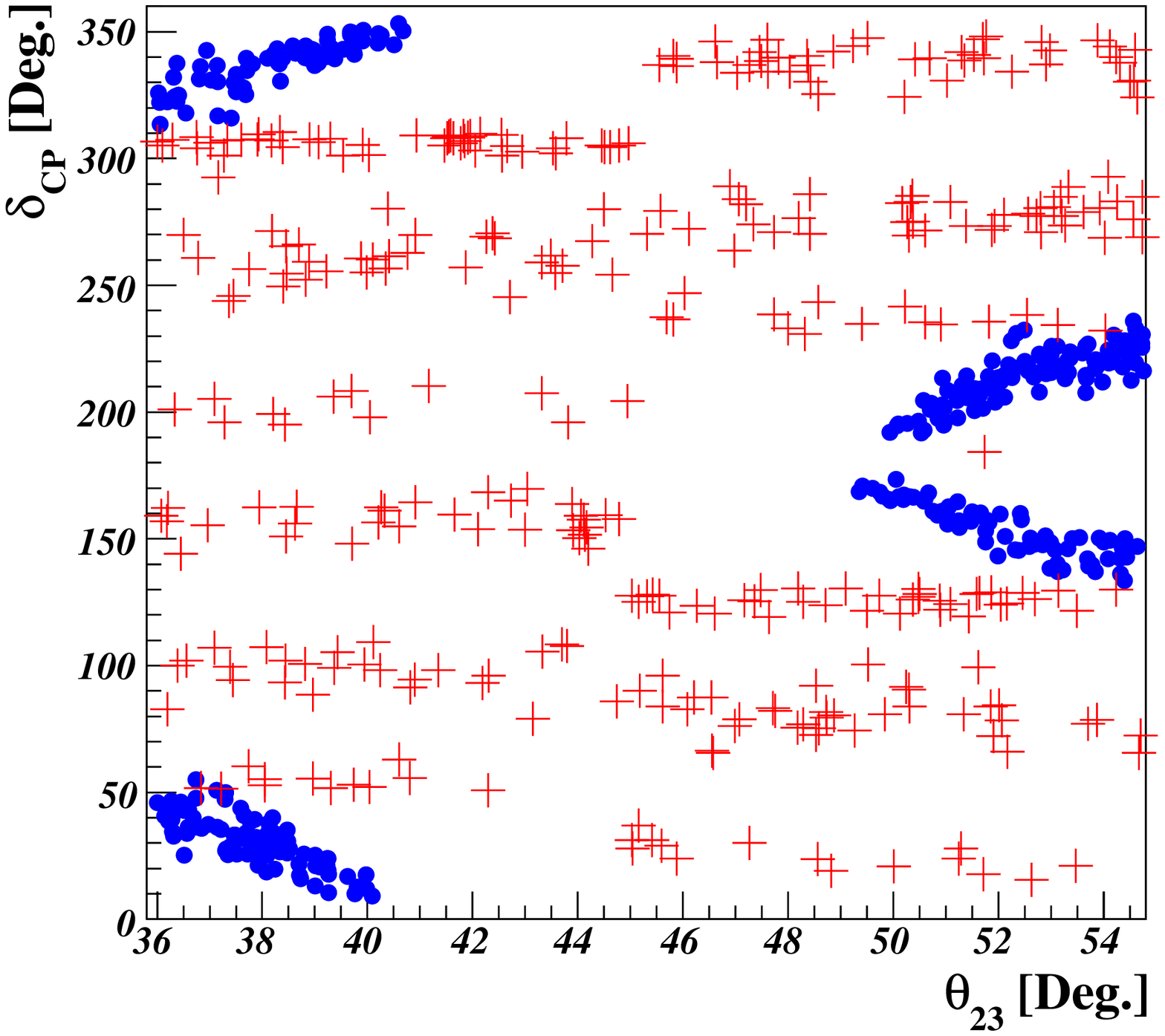,width=6.5cm,angle=0}
\end{minipage}
\caption{\label{FigA1pp}
The Jarlskog invariant $J_{CP}$ versus the reactor angle $\theta_{13}$ (left plot), and the Dirac $CP$ phase $\delta_{CP}$ versus $\theta_{23}$ (right plot). The (red) crosses and (blue) dots represent the results for the inverted and normal mass hierarchy, respectively. The vertical dotted (solid) lines indicate the upper and lower bounds on $\theta_{13}$ given in Eq.~(\ref{expnu}) at the $3\sigma$ level (best-fit value).}
\end{figure}

\begin{figure}[t]
\begin{minipage}[t]{6.0cm}
\epsfig{figure=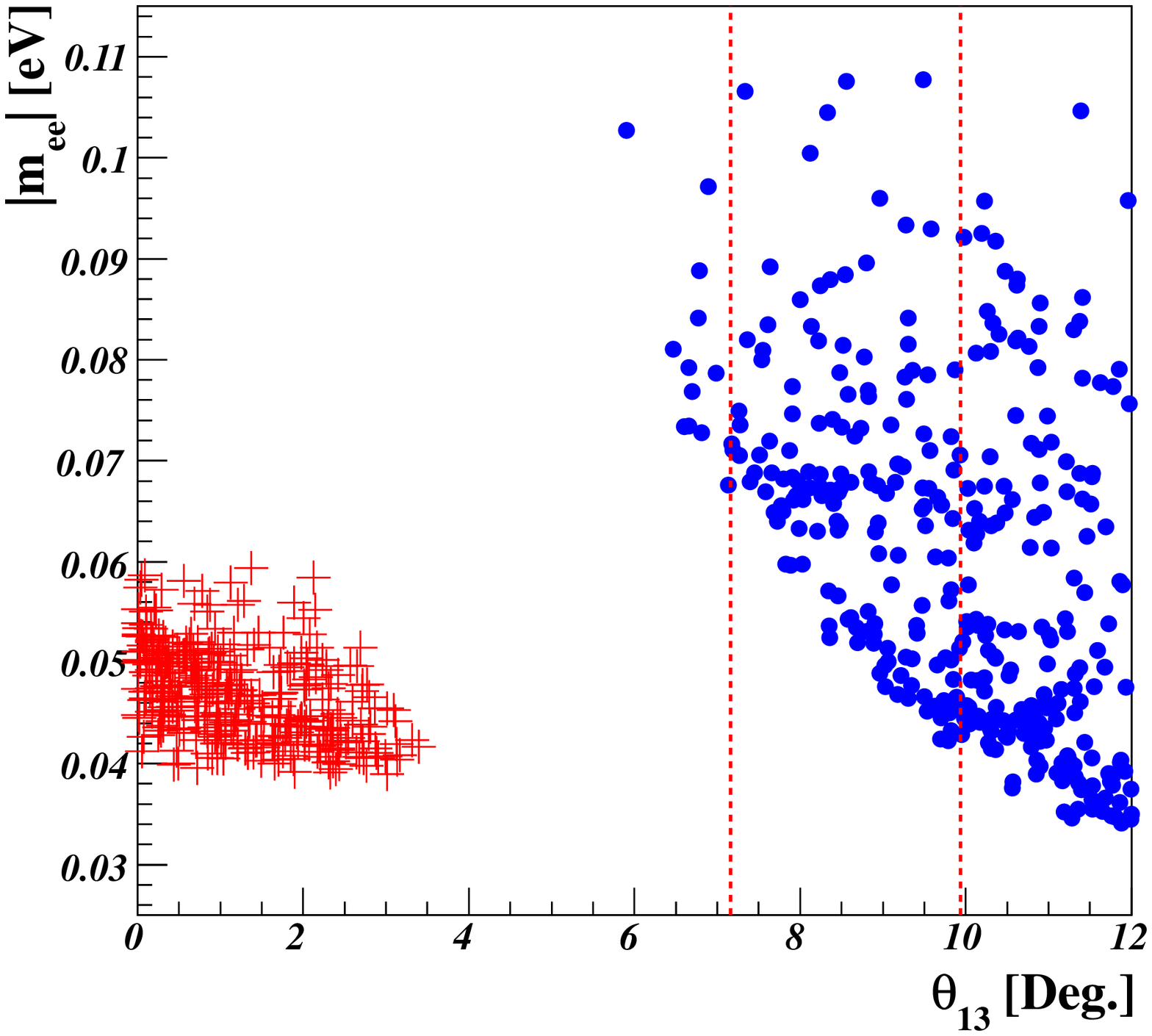,width=6.5cm,angle=0}
\end{minipage}
\hspace*{1.0cm}
\begin{minipage}[t]{6.0cm}
\epsfig{figure=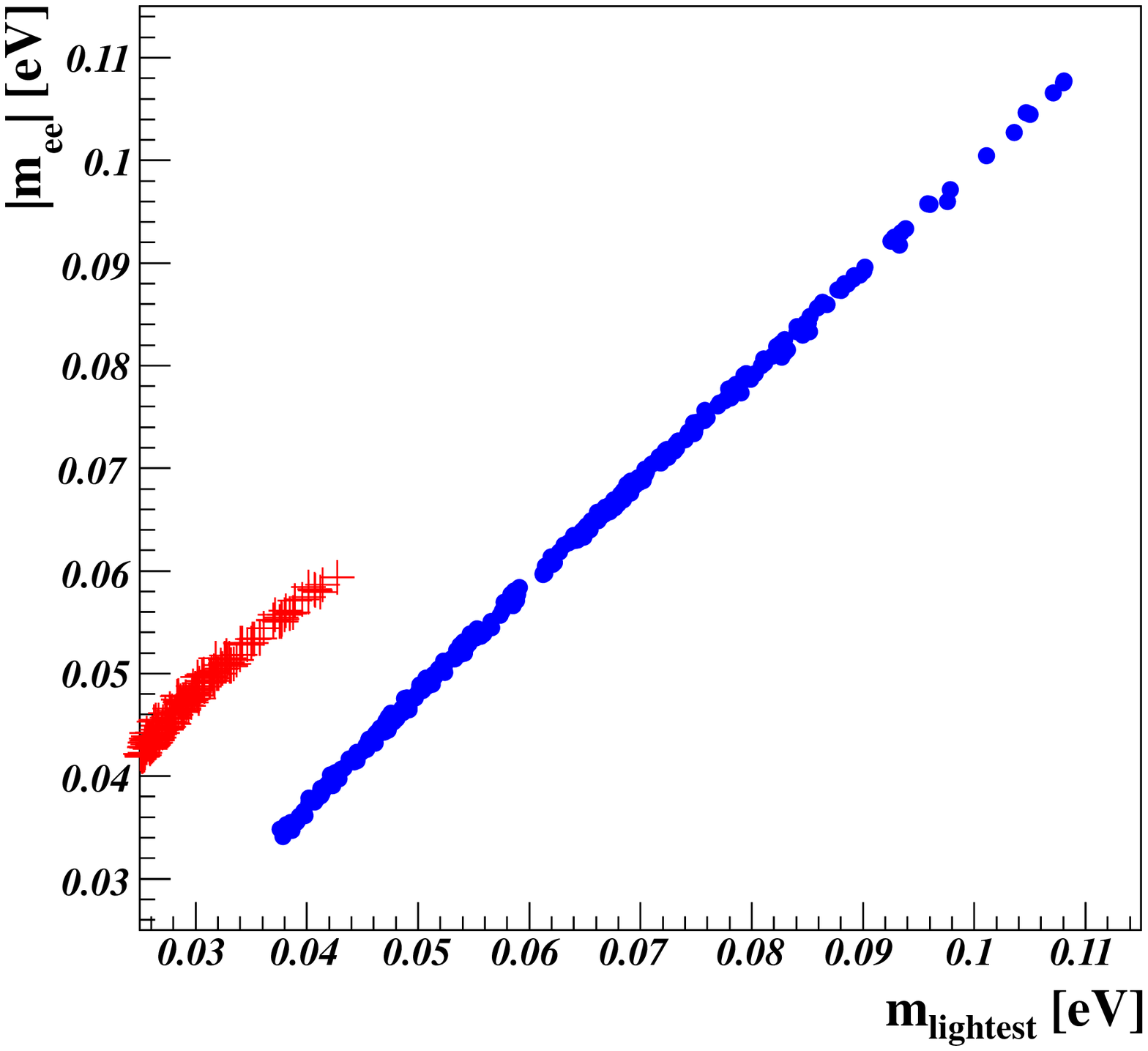,width=6.5cm,angle=0}
\end{minipage}
\caption{\label{FigA3} Plots of $|m_{ee}|$ as a function of $\theta_{13}$ and $m_{\rm lightest}$. The red crosses and the blue dots represent results for the inverted and the normal mass hierarchy, respectively. The vertical dashed lines show the experimental bounds of Eq.~(\ref{expnu}) at $3\sigma$'s.}
\end{figure}

Using the formulae for the neutrino mixing angles and masses and our values of $M,v_\Phi$, we obtain the following allowed regions of the unknown model parameters: for the normal mass hierarchy (NMH),
 \begin{eqnarray}
  &&0.21\lesssim\kappa\lesssim0.86~,\qquad1.04\lesssim y_{1}\lesssim1.37~,\qquad1.04\lesssim y_{2}\lesssim1.39~,\nonumber\\
  &&96^{\circ}\lesssim\varphi\lesssim121^{\circ}~~{\rm and~}~239^{\circ}\lesssim\varphi\lesssim257^{\circ}~,\qquad
1.1\lesssim m_{0}\times10^{-2}{\rm [eV]}\lesssim3.6~;
  \label{input1}
 \end{eqnarray}
for the inverted mass hierarchy (IMH),
 \begin{eqnarray}
  &&0.81\lesssim\kappa\lesssim2~,\qquad0.91\lesssim y_{1}\lesssim1.09~,\qquad0.91\lesssim y_{2}\lesssim1.08~,\nonumber\\
  && 116^{\circ}\lesssim\varphi\lesssim248^{\circ}
  ~,\qquad\qquad1.6\lesssim m_{0}\times10^{-2}{\rm [eV]}\lesssim2.2~.
  \label{input2}
 \end{eqnarray}
\begin{figure}[t]
\begin{minipage}[t]{6.0cm}
\epsfig{figure=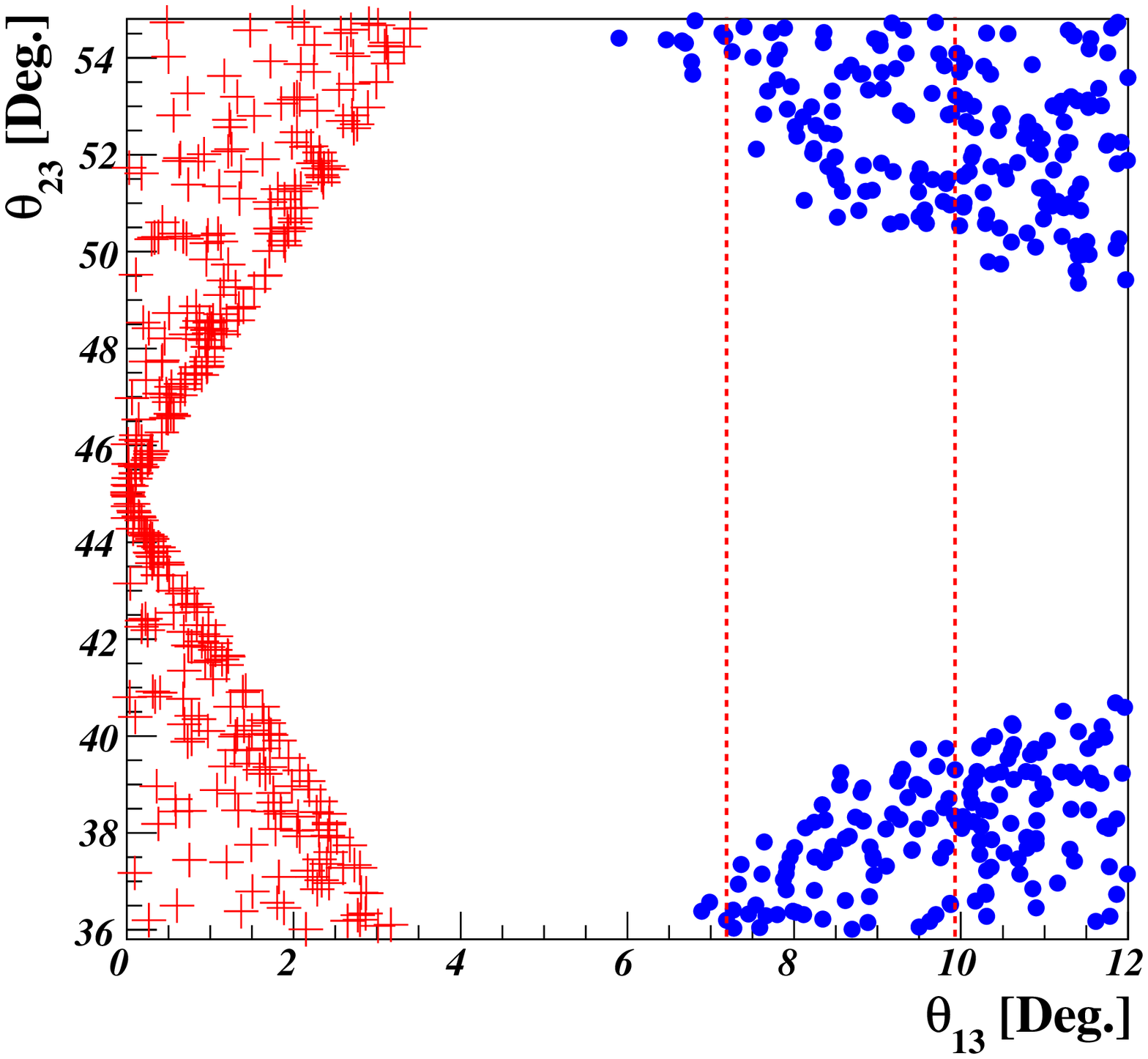,width=6.5cm,angle=0}
\end{minipage}
\hspace*{1.0cm}
\begin{minipage}[t]{6.0cm}
\epsfig{figure=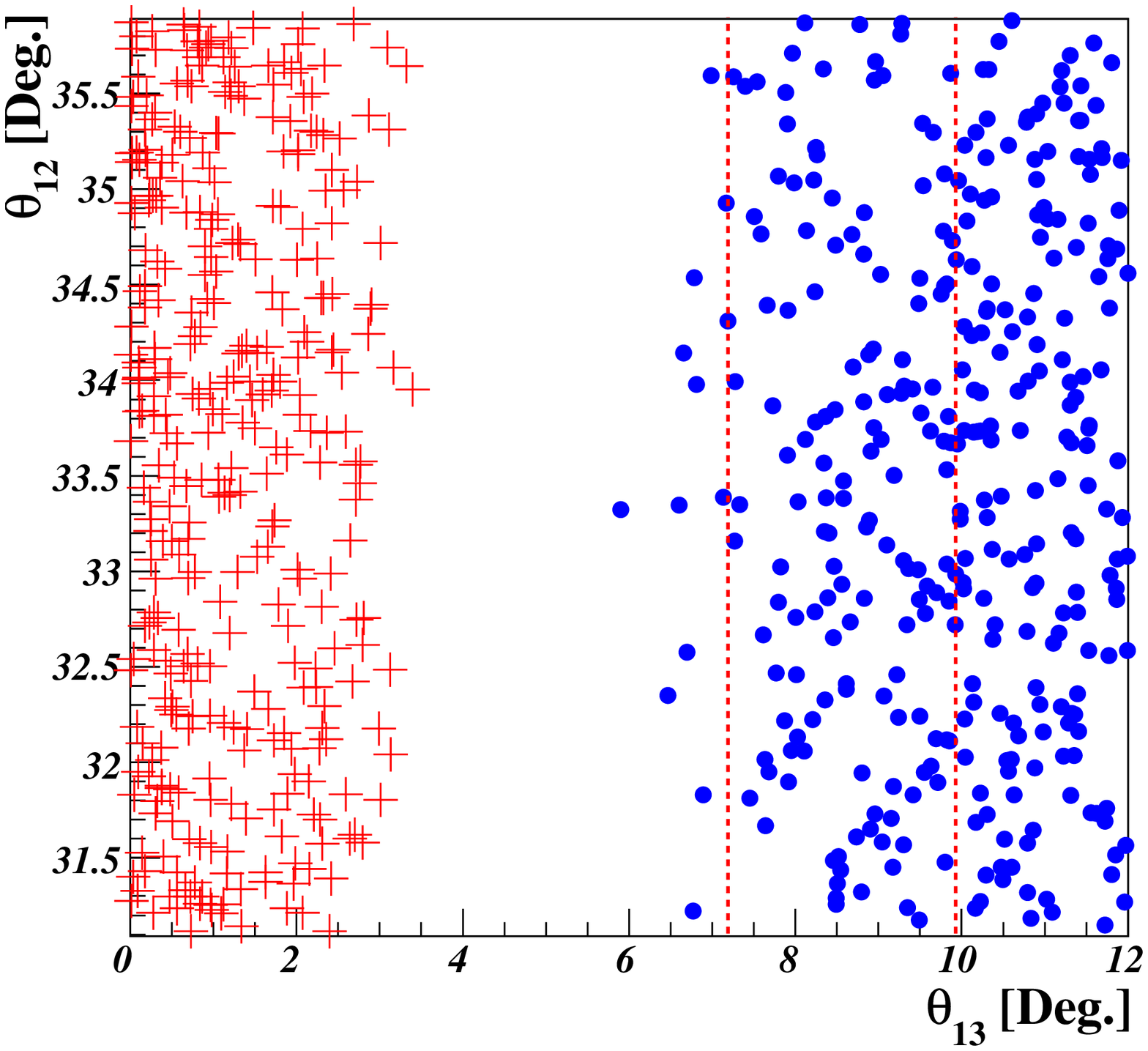,width=6.5cm,angle=0}
\end{minipage}
\caption{\label{FigA2} The behaviors of $\theta_{23}$ and $\theta_{12}$ in terms of  $\theta_{13}$. The red crosses and the blue dots represent results for the inverted mass hierarchy and the normal mass hierarchy, respectively. The dotted vertical lines represent the experimental bounds of Eq.~(\ref{expnu}) at $3\sigma$'s.
}
\end{figure}
Note that here we have used the $3\sigma$ experimental bounds on $\theta_{12},\theta_{23},\Delta m^{2}_{\rm Sol}, \Delta m^{2}_{\rm Atm}$ in Eq.~(\ref{expnu}), except for $\theta_{13}<12^{\circ}$ for which we use the values in Eqs.~(\ref{input1},\ref{input2}).
For these parameter regions, we investigate how mixing parameters do behave for the NMH and
IMH. In Figs.~\ref{FigA1}-\ref{FigA3}, the data points represented by blue dots and red crosses indicate results for the NMH and IMH, respectively. The left-hand-side plot in Fig.~\ref{FigA1} shows how the mixing angle $\theta_{13}$ depends on the ratio $y_1/y_2=y_1^{\nu}/y_2^{\nu}$ of the third- and second-generation neutrino Yukawa couplings; the right-hand-side plot shows how $\theta_{13}$ depends on the parameter $\kappa=y^{\nu}_R v_\chi/M$.
Fig.~\ref{FigA1p} shows the mixing angle $\theta_{13}$ as a function of the phase $\varphi$ of $y^{\nu}_R v_\chi/M$.
As can be seen in Figs.~\ref{FigA1}-\ref{FigA1p}, only normal mass hierarchy is permitted within $3\sigma$ experimental bounds. And we see that the measured value of $\theta_{13}$ from the Daya Bay and RENO experiments can be achieved at $3\sigma$'s for $0.92\lesssim y_{1}/y_{2}<1$, $1<y_{1}/y_{2}\lesssim1.06$, $0.2\lesssim\kappa\lesssim0.7$, $95^{\circ}\lesssim\varphi\lesssim115^{\circ}$ and $245^{\circ}\lesssim\varphi\lesssim255^{\circ}$ for NMH. For IMH, in Figs.~\ref{FigA1}-\ref{FigA1p} the value of $\theta_{13}$ reaches at most $3^{\circ}$, which is excluded by the measurements of $\theta_{13}$.

Fig.~\ref{FigA2} shows how the values of $\theta_{13}$ depend on the mixing angles $\theta_{23}$ and $\theta_{12}$. As can be seen in the left plot of Fig.~\ref{FigA2}, the behavior of $\theta_{23}$ in terms of the measured values of $\theta_{13}$ at $3\sigma$'s for the NMH is different than for the IMH. As already mentioned, the IMH is excluded by the measured values of $\theta_{13}$ in Fig.~\ref{FigA2}. For the NMH we see that the measured values of $\theta_{13}$ can be achieved for $49.5^{\circ}\lesssim\theta_{23}\lesssim54.8^{\circ}$ and $35.8^{\circ}\lesssim\theta_{23}\lesssim40.5^{\circ}$, with large deviations from maximality, which are favored at $1\sigma$ by the experimental bounds as can be seen in Eq.~(\ref{expnu}).
Future precise measurements of $\theta_{23}$, whether $\theta_{23}\rightarrow45^{\circ}$ or $|\theta_{23}-45^{\circ}|\rightarrow5^{\circ}$, will either exclude or favor our model.
From the right plot of Fig.~\ref{FigA2}, we see that the
predictions for $\theta_{13}$ do not strongly depend on $\theta_{12}$ in the allowed region.

To see how the parameters are correlated with low-energy $CP$ violation observables measurable through neutrino oscillations, we consider the leptonic $CP$ violation parameter defined by the
Jarlskog invariant in Eq.~(\ref{JCP}) which can be expressed in terms of the elements of the matrix $h=\mathcal{M}_{\nu}\mathcal{M}^{\dag}_{\nu}$~\cite{Branco:2002xf}:
 \begin{eqnarray}
  J_{CP}=-\frac{{\rm Im}\{h_{12}h_{23}h_{31}\}}{\Delta m^{2}_{21}\Delta m^{2}_{31}\Delta m^{2}_{32}}~.
  \label{JCPA}
 \end{eqnarray}
The behavior of $J_{CP}$ as a function of $\theta_{13}$ is plotted on the left plot of Fig.~\ref{FigA1pp}.
We see that the value of $J_{CP}$ lies in the ranges $0.006\sim0.03$ and $-0.03\sim-0.004$ (NMH) for the measured value of $\theta_{13}$ at $3\sigma$'s. Also, in our model we have
 \begin{eqnarray}
  {\rm Im}\{h_{12}h_{23}h_{31}\}=\frac{27m^{6}_{0}}{4a^{4}b^{3}}y^{2}_{1}y^{2}_{2}(y^{2}_{1}-y^{2}_{2})\sin\psi_{2}\{....\}~,
  \label{JCPB}
 \end{eqnarray}
in which $\{.....\}$ stands for a complicated lengthy function of $y_{1}$, $y_{2}$, $a$, $b$, $\psi_{1}$ and $\psi_{2}$. Clearly, Eq.~(\ref{JCPB}) indicates that in the limit of $y_{2}\rightarrow y_{1}$ or $\sin\psi_{2}\rightarrow0$ the leptonic $CP$ violation $J_{CP}$ goes to zero.
 When $y_{2}\neq y_{1}$, i.e.\ for the IMH case, $J_{CP}$ could go to zero as $\sin\psi_{2}$ of Eq.~(\ref{JCPB}) [see, Eq.~(\ref{alphs_beta}) and Fig.~\ref{FigA1p}].
In the case of the NMH, $J_{CP}$ has nonzero values for the measured range of $\theta_{13}$ while $J_{CP}$ goes to zero for $\theta_{13}\rightarrow0$, which corresponds to $y_{2}\rightarrow y_{1}$.
The right plot of Fig.~\ref{FigA1pp} shows the behavior of the Dirac $CP$ phase $\delta_{CP}$ [see Eq.~(\ref{DeltaCP1})] as a function of $\theta_{23}$, where the values of $\delta_{CP}$ lie in the ranges $0^{\circ}<\delta_{CP}\lesssim60^{\circ}, 130^{\circ}\lesssim\delta_{CP}<180^{\circ}, 180^{\circ}<\delta_{CP}\lesssim240^{\circ}$ and $310^{\circ}\lesssim\delta_{CP}<360^{\circ}$ for the NMH (for the IMH, $\delta_{CP}$ can vary over a wide range, but which is excluded by the measured values of $\theta_{13}$). Interestingly, for the best-fit values of $\theta_{23}$ the values of $\delta_{CP}$ are predicted as the one around $10^{\circ}, 170^{\circ}, 190^{\circ}, 350^{\circ}$ for NMH. So, future precise measurements of $\theta_{23}$ will provide more information on $\delta_{CP}$.

Moreover, we can straightforwardly obtain the effective neutrino mass $|m_{ee}|$ that characterizes the amplitude for neutrinoless double beta decay~\cite{Bilenky:2010zz}:
 \begin{eqnarray}
  |m_{ee}|\equiv \left|\sum_{i}(U_{\rm PMNS})^{2}_{ei}m_{i}\right|~,
  \label{mee}
 \end{eqnarray}
where $U_{\rm PMNS}$ is given in a good approximation as Eq.~(\ref{Unu}).
The left and right plots in Fig.~\ref{FigA3} show the behavior of the effective neutrino mass $|m_{ee}|$ in terms of $\theta_{13}$ and the lightest neutrino mass, respectively.
In the left plot of Fig.~\ref{FigA3}, for the measured values of $\theta_{13}$ at $3\sigma$'s, the effective neutrino mass $|m_{ee}|$ can be in the range $0.04\lesssim|m_{ee}|[{\rm eV}]<0.11$ for NMH.
The right plot of Fig.~\ref{FigA3} shows $|m_{ee}|$ as a function of $m_{\rm lightest}$, where $m_{\rm lightest}=m_{1}$ for the NMH and $m_{\rm lightest}=m_{3}$ for the IMH.
Our model predicts that the effective mass $|m_{ee}|$ is within the sensitivity about $10^{-2}$ eV of planned neutrinoless double-beta decay experiments~\cite{Aalseth:2004hb}.

\section{Conclusion}
Under $SU(2)_{L}\times U(1)_{Y}$ gauge symmetry, we have proposed a new model of leptons and quarks based on the discrete flavor symmetry $T'$, the double covering of $A_4$. Here we impose that all Yukawa couplings be of order one, which implies that the hierarchies of charged fermion masses and the mildness of neutrino masses are responsible for six types of Higgs scalars.
In addition to the gauge and flavor symmetries, in order to simplify our model and to remove the unwanted Yukawa terms appearing in the Lagrangian we have introduced a continuous global symmetry $U(1)_X$ which can not be gauged. After spontaneous $U(1)_X$ breaking, to avoid Goldstone bosons it has to be explicitly broken down to a subgroup $Z_2$. After spontaneous breaking of flavor symmetry, with the constraint of renormalizability in the Lagrangian, the leptons have $m_{e}=0$ and the quarks have CKM mixing angles $\theta^{q}_{12}=13^{\circ}, \theta^{q}_{23}=0^{\circ}$ and $\theta^{q}_{13}=0^{\circ}$. Thus, certain effective dimension-5 operators driven by the gauge-singlet and $T'$-triplet $\chi$ field are introduced as an equal footing, which induce $m_{e}\neq0$ and lead the quark mixing matrix to the CKM one in form. On the other hand, the neutrino Lagrangian still keeps renormalizability. We have assumed that there is a cutoff scale $\Lambda$, above which there exists unknown physics.

We have shown numerical analysis in the lepton sector of our model, where only normal mass hierarchy is permitted within $3\sigma$ experimental bounds with the prediction of both large deviations from maximality in the atmospheric mixing angle $\theta_{23}$ and the measured values of reactor angle. So, future precise measurements of $\theta_{23}$, whether $\theta_{23}\rightarrow45^{\circ}$ or $|\theta_{23}-45^{\circ}|\rightarrow5^{\circ}$, would either exclude or favor our model. Together with it, our model has made predictions both for the Dirac $CP$ phase $0^{\circ}<\delta_{CP}\lesssim60^{\circ}, 130^{\circ}\lesssim\delta_{CP}<180^{\circ}, 180^{\circ}<\delta_{CP}\lesssim240^{\circ}$ and $310^{\circ}\lesssim\delta_{CP}<360^{\circ}$, which is almost compatible with the global analysis in $1\sigma$ experimental bounds. Moreover, we have shown the effective mass $|m_{ee}|$ measurable in neutrinoless double beta decay to be in the range $0.04\lesssim|m_{ee}|[eV]<0.11$ for the normal hierarchy, which could be tested in near future neutrino experiments.

\newpage
\appendix
\section{The Higgs potential}
\label{AHiggs}
\vspace{0.5cm}
\begin{figure}[h]
  \epsfig{figure=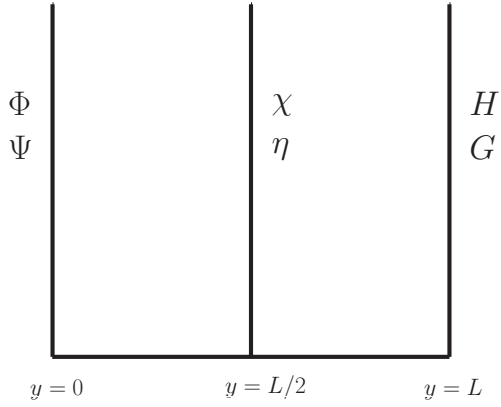,width=7cm,angle=0}
 \caption{\label{fig:exd}
  Fifth dimension and locations of scalar and fermion fields.}
\end{figure}
In this Appendix, as an example, we present our Higgs potential and its minimization, as well as our prescription for effecting the stability of the vacuum alignment.
We solve the vacuum alignment problem by extending the model into a spatial extra dimension
$y$~\cite{Altarelli:2005yp}. We assume that each field lives on a 4D brane either at $y = 0$ or  at $y = L/2$ or at
$y = L$, as shown in Fig.~\ref{fig:exd}. The heavy neutrino masses and the neutrino Yukawa interactions arise from local operators at the branes $y=0$ and $y = L/2$, while the charged fermion masses are realized by non-local effects involving both branes $y = 0$, $y = L/2$ and $y = L$. We impose that all the parameters appearing in the Lagrangian are assumed to be real. Once the scalars $\Phi, \Psi, \chi, H, G$ acquire complex VEVs at the different branes, the $CP$ symmetry can be spontaneously broken. A rigorous explanation of this possibility is beyond the scope of this paper.

The most general scalar potential ($d\leq5$) for the Higgs fields $\Phi,H,\Psi,G, \eta$ and $\chi$ invariant under $SU(2)_{L}\times U(1)_{Y}\times T'$ and obeying the conditions in the previous paragraph, is then given by
\begin{eqnarray}
V & =&  V_{y=0}+V_{y=\frac{L}{2}}+V_{y=L} ,
\label{potential}
\end{eqnarray}
where
\begin{eqnarray}
V_{y=0} &=& V(\Phi)+V(\Psi)+V(\Phi\Psi)\nonumber\\
V_{y=\frac{L}{2}}&=&V(\chi)+V(\eta)+V(\chi\eta)\nonumber\\
V_{y=L}  &=&V(H)+V(G)+V(HG) ~,
\end{eqnarray}
and~\footnote{In $V(H)$ and $V(G)$ the terms $i\mu^{2}_{H}(H^{\dag}H)_{\mathbf{1}}$ and $i\mu^{2}_{G}(G^{\dag}G)_{\mathbf{1}}$ are expanded as $i\mu^{2}_{H}(H^{\dag}_{1}H_2-H^{\dag}_{2}H_1)$ and $i\mu^{2}_{G}(G^{\dag}_{1}G_2-G^{\dag}_{2}G_1)$, respectively.}
 \begin{eqnarray}
V(\Phi) &=& \label{pot1} \mu^{2}_{\Phi}(\Phi^{\dag}\Phi)_{\mathbf{1}}+\lambda^{\Phi}_{1}(\Phi^{\dag}\Phi)_{\mathbf{1}}(\Phi^{\dag}\Phi)_{\mathbf{1}}+\lambda^{\Phi}_{2}
            (\Phi^{\dag}\Phi)_{\mathbf{1^\prime}}(\Phi^{\dag}\Phi)_{\mathbf{1^{\prime\prime}}}\nonumber\\
&+&\lambda^{\Phi}_{3}(\Phi^{\dag}\Phi)_{\mathbf{3}_{s}}(\Phi^{\dag}\Phi)_{\mathbf{3}_{s}}
  +\lambda^{\Phi}_{4}(\Phi^{\dag}\Phi)_{\mathbf{3}_{a}}(\Phi^{\dag}\Phi)_{\mathbf{3}_{a}}+\lambda^{\Phi}_{5}\{(\Phi^{\dag}\Phi)_{\mathbf{3}_{s}}(\Phi^{\dag}\Phi)_{\mathbf{3}_{a}}+h.c.\}~,\\
V(\Psi) &=&  \mu^{2}_{\Psi}(\Psi^{\dag}\Psi)_{\mathbf{1}}+\lambda^{\Psi}_{1}(\Psi^{\dag}\Psi)_{\mathbf{1}}(\Psi^{\dag}\Psi)_{\mathbf{1}}+\lambda^{\Psi}_{2}
            (\Psi^{\dag}\Psi)_{\mathbf{1^\prime}}(\Psi^{\dag}\Psi)_{\mathbf{1^{\prime\prime}}}
+\lambda^{\Psi}_{3}(\Psi^{\dag}\Psi)_{\mathbf{3}_{s}}(\Psi^{\dag}\Psi)_{\mathbf{3}_{s}}\nonumber\\
  &+&\lambda^{\Psi}_{4}(\Psi^{\dag}\Psi)_{\mathbf{3}_{a}}(\Psi^{\dag}\Psi)_{\mathbf{3}_{a}}+\lambda^{\Psi}_{5}\{(\Psi^{\dag}\Psi)_{\mathbf{3}_{s}}(\Psi^{\dag}\Psi)_{\mathbf{3}_{a}}+h.c.\}~,\\
V(\Phi\Psi) &=& \label{pot11} \lambda^{\Phi\Psi}_{1}(\Phi^{\dag}\Phi)_{\mathbf{1}}(\Psi^{\dag}\Psi)_{\mathbf{1}}+\lambda^{\Phi\Psi}_{2}
            (\Phi^{\dag}\Phi)_{\mathbf{1^\prime}}(\Psi^{\dag}\Psi)_{\mathbf{1^{\prime\prime}}}+\lambda^{\Phi\Psi}_{3}(\Phi^{\dag}\Phi)_{\mathbf{3}_{s}}(\Psi^{\dag}\Psi)_{\mathbf{3}_s}\nonumber\\
  &+&\lambda^{\Phi\Psi}_{4}(\Phi^{\dag}\Phi)_{\mathbf{3}_{a}}(\Psi^{\dag}\Psi)_{\mathbf{3}_a}+\lambda^{\Phi\Psi}_{5}\{(\Phi^{\dag}\Phi)_{\mathbf{3}_{s}}(\Psi^{\dag}\Psi)_{\mathbf{3}_a}+h.c.\}\nonumber\\
  &+&\lambda^{\Phi\Psi}_{6}\{(\Phi^{\dag}\Phi)_{\mathbf{3}_{a}}(\Psi^{\dag}\Psi)_{\mathbf{3}_s}+h.c.\}
+\lambda^{\Phi\Psi}_{7}\left\{(\Phi^{\dag}\Psi)_{\mathbf{1}}(\Phi^{\dag}\Psi)_{\mathbf{1}}+h.c.\right\}\nonumber\\
&+&\lambda^{\Phi\Psi}_{8}
            \left\{(\Phi^{\dag}\Psi)_{\mathbf{1^\prime}}(\Phi^{\dag}\Psi)_{\mathbf{1^{\prime\prime}}}+h.c.\right\}+\lambda^{\Phi\Psi}_{9}\left\{(\Phi^{\dag}\Psi)_{\mathbf{3}_{s}}(\Phi^{\dag}\Psi)_{\mathbf{3}_s}+h.c.\right\}\nonumber\\
  &+&\lambda^{\Phi\Psi}_{10}\left\{(\Phi^{\dag}\Psi)_{\mathbf{3}_{a}}(\Phi^{\dag}\Psi)_{\mathbf{3}_a}+h.c.\right\}+\lambda^{\Phi\Psi}_{11}\left\{(\Phi^{\dag}\Psi)_{\mathbf{3}_{s}}(\Phi^{\dag}\Psi)_{\mathbf{3}_a}+h.c.\right\}\nonumber\\
  &+&\lambda^{\Phi\Psi}_{12}(\Phi^{\dag}\Psi)_{\mathbf{1}}(\Psi^{\dag}\Phi)_{\mathbf{1}}+\lambda^{\Phi\Psi}_{13}\left\{
            (\Phi^{\dag}\Psi)_{\mathbf{1^\prime}}(\Psi^{\dag}\Phi)_{\mathbf{1^{\prime\prime}}}+h.c.\right\}+\lambda^{\Phi\Psi}_{14}(\Phi^{\dag}\Psi)_{\mathbf{3}_{s}}(\Psi^{\dag}\Phi)_{\mathbf{3}_s}\nonumber\\
&+&\lambda^{\Phi\Psi}_{15}(\Phi^{\dag}\Psi)_{\mathbf{3}_{a}}(\Psi^{\dag}\Phi)_{\mathbf{3}_a}+\lambda^{\Phi\Psi}_{16}\{(\Phi^{\dag}\Psi)_{\mathbf{3}_{s}}(\Psi^{\dag}\Phi)_{\mathbf{3}_a}+h.c.\}~,
 \end{eqnarray}
 \begin{eqnarray}
V(H) &=& i\mu^{2}_{H}(H^{\dag}H)_{\mathbf{1}}+ \lambda^{H}_{1}(H^{\dag}H)_{\mathbf{1}}(H^{\dag}H)_{\mathbf{1}}+\lambda^{H}_{2}\{(H^{\dag}H)_{\mathbf{3}}(H^{\dag}H)_{\mathbf{3}}+h.c.\}~,\\
V(G) &=& i\mu^{2}_{G}(G^{\dag}G)_{\mathbf{1}}+ \lambda^{G}_{1}(G^{\dag}G)_{\mathbf{1}}(G^{\dag}G)_{\mathbf{1}}+\lambda^{G}_{2}\{(G^{\dag}G)_{\mathbf{3}}(G^{\dag}G)_{\mathbf{3}}+h.c.\}~,\\
V(H G) &=& \lambda^{H G}_{1}(H^{\dag}H)_{\mathbf{1}}(G^{\dag}G)_{\mathbf{1}}+\lambda^{H G}_{2}(H^{\dag}G)_{\mathbf{1}}(G^{\dag}H)_{\mathbf{1}}+\lambda^{H G}_{3}\{(G^{\dag}H)_{\mathbf{1}}(G^{\dag}H)_{\mathbf{1}}+h.c.\}\nonumber\\
&+&\lambda^{H G}_{4}\{(H^{\dag}H)_{\mathbf{3}}(G^{\dag}G)_{\mathbf{3}}+h.c.\}+\lambda^{H G}_{5}\{(H^{\dag}G)_{\mathbf{3}}(G^{\dag}H)_{\mathbf{3}}+h.c.\}\nonumber\\
&+&\lambda^{H G}_{6}\{(G^{\dag}H)_{\mathbf{3}}(G^{\dag}H)_{\mathbf{3}}+h.c.\}~,
 \end{eqnarray}
 \begin{eqnarray}
V(\chi) &=& \mu^{2}_{\chi}\left\{(\chi\chi)_{\mathbf{1}}+(\chi^{\ast}\chi^{\ast})_{\mathbf{1}}\right\}+m^{2}_{\chi}(\chi\chi^{\ast})_{\mathbf{1}}+\lambda^{\chi}_{1}\left\{(\chi\chi)_{\mathbf{1}}(\chi\chi)_{\mathbf{1}}+(\chi^{\ast}\chi^{\ast})_{\mathbf{1}}(\chi^{\ast}\chi^{\ast})_{\mathbf{1}}\right\}\nonumber\\
&+&\lambda^{\chi}_{2}
            \left\{(\chi\chi)_{\mathbf{1}^\prime}(\chi\chi)_{\mathbf{1}^{\prime\prime}}+(\chi^{\ast}\chi^{\ast})_{\mathbf{1}^\prime}(\chi^{\ast}\chi^{\ast})_{\mathbf{1}^{\prime\prime}}\right\}\nonumber\\
&+&\tilde{\lambda}^{\chi}_{2} \left\{(\chi^{\ast}\chi)_{\mathbf{1}^\prime}(\chi\chi)_{\mathbf{1}^{\prime\prime}}+(\chi^{\ast}\chi)_{\mathbf{1}^{\prime\prime}}(\chi^{\ast}\chi^{\ast})_{\mathbf{1}^{\prime}}\right\}\nonumber\\
  &+&\lambda^{\chi}_{3}\left\{(\chi\chi)_{\mathbf{3}_{s}}(\chi\chi)_{\mathbf{3}_{s}}+(\chi^{\ast}\chi^{\ast})_{\mathbf{3}_{s}}(\chi^{\ast}\chi^{\ast})_{\mathbf{3}_{s}}\right\}+\tilde{\lambda}^{\chi}_{3}(\chi^{\ast}\chi)_{\mathbf{3}_{s}}\left\{(\chi\chi)_{\mathbf{3}_{s}}+(\chi^{\ast}\chi^{\ast})_{\mathbf{3}_{s}}\right\}\nonumber\\
  &+&\lambda^{\chi}_{4}\left\{(\chi^\ast\chi)_{\mathbf{3}_{a}}(\chi\chi)_{\mathbf{3}_{s}}+(\chi\chi^{\ast})_{\mathbf{3}_{a}}(\chi^{\ast}\chi^{\ast})_{\mathbf{3}_{s}}\right\}\nonumber\\
  &+&\xi^{\chi}_{1}\left\{\chi(\chi\chi)_{\mathbf{3}_{s}}+\chi^{\ast}(\chi^{\ast}\chi^{\ast})_{\mathbf{3}_{s}}\right\}+\tilde{\xi}^{\chi}_{1}\left\{\chi(\chi^{\ast}\chi^{\ast})_{\mathbf{3}_{s}}+\chi^{\ast}(\chi\chi)_{\mathbf{3}_{s}}\right\}~,\nonumber\\
&+&\frac{\zeta^{\chi}_{1}}{\Lambda}\left\{(\chi\chi)_{\mathbf{1}}(\chi\chi\chi)_{\mathbf{1}}+(\chi^{\ast}\chi^{\ast})_{\mathbf{1}}(\chi^{\ast}\chi^{\ast}\chi^{\ast})_{\mathbf{1}}\right\}+...+\frac{\zeta^{\chi}_{13}}{\Lambda}\{...\}~,
\label{pot2}
 \end{eqnarray}
 \begin{eqnarray}
V(\eta) &=&\label{pot21} \mu^{2}_{\eta}(\eta^{\dag}\eta)+\lambda^{\eta}(\eta^{\dag}\eta)^{2}~,\\
V(\chi\eta) &=& \label{pot22} \lambda^{\chi\eta}_{1}(\eta^{\dag}\eta)\{(\chi\chi)_{\mathbf{1}}+(\chi^{\ast}\chi^{\ast})_{\mathbf{1}}\}\nonumber\\
&+&\frac{\xi^{\chi\eta}_{1}}{\Lambda}(\eta^{\dag}\eta)\{(\chi\chi\chi)_{\mathbf{1}}+(\chi^{\ast}\chi^{\ast}\chi^{\ast})_{\mathbf{1}}\}+\frac{\xi^{\chi\eta}_{2}}{\Lambda}(\eta^{\dag}\eta)\{(\chi^{\ast}\chi\chi)_{\mathbf{1}}+(\chi\chi^{\ast}\chi^{\ast})_{\mathbf{1}}\}~.
 \end{eqnarray}
Here, $\mu_{\Phi},\mu_{\Psi},\mu_{H},\mu_{G},\mu_{\eta},\mu_{\chi}$, $m_{\chi}$, $\xi^{\chi}_{1}$,  $\tilde{\xi}^{\chi}_{1}$, $\zeta^{\chi}_{1...13}$ and $\xi^{\chi\eta}_{1,2}$ have a mass dimension,
whereas $\lambda^{\Phi}_{1,...,5}$, $\lambda^{\Psi}_{1,...,5}$, $\lambda^{H}_{1,2}$, $\lambda^{G}_{1,2}$, $\lambda^{\eta}$, $\lambda^{\chi}_{1,...,4}$, $\tilde{\lambda}^{\chi}_{2,3}$, $\lambda^{\Phi\Psi}_{1,...,16}$, $\lambda^{H G}_{1,...,4}$ and $\lambda^{\chi\eta}$ are all dimensionless. And in $V(\chi)$ $``\cdots"$ denotes dimension-5 operators composed of all possible combinations of $\chi$ fields.

\subsection{Minimization of the neutral scalar potential}
After the breaking of the flavor and electroweak symmetry, in the neutral Higgs sector, in order to find minimum configuration of the Higgs potential, we in general let
 \begin{eqnarray}
  \langle\Phi_{i}\rangle &=&
  {\left(\begin{array}{c}
  0 \\
  \frac{1}{\sqrt{2}}v_{\Phi_j}e^{i\gamma_{j}}
 \end{array}\right)}~,\quad
  \langle\Psi_{j}\rangle =
  {\left(\begin{array}{c}
  0 \\
  \frac{1}{\sqrt{2}}v_{\Psi_j}e^{i\zeta_{j}}
 \end{array}\right)}~,  \qquad\langle\eta\rangle =
  {\left(\begin{array}{c}
  0 \\
  \frac{1}{\sqrt{2}}v_{\eta}e^{i\theta}
 \end{array}\right)}~,\nonumber\\
  \langle H_{k}\rangle &=&
  {\left(\begin{array}{c}
  0 \\
  \frac{1}{\sqrt{2}}v_{H_k}e^{i\rho_k}
 \end{array}\right)}~,\quad
  \langle G_{k}\rangle =
  {\left(\begin{array}{c}
  0 \\
  \frac{1}{\sqrt{2}}v_{G_k}e^{i\sigma_k}
 \end{array}\right)}~, ~\quad\langle\chi_{j}\rangle= v_{\chi_{j}}e^{i\varphi_{j}}~,
  \label{neuvevs}
 \end{eqnarray}
with $j=1-3, k=1,2$, where $v_{\Phi_j}, v_{\Psi_j}, v_{\eta}, v_{H_k}, v_{G_k} v_{\chi_{j}}$ are real and positive, and $\gamma_{j}$, $\zeta_{j}$, $\rho_{k}, \sigma_{k}, \varphi_{j}$ are physically meaningful phases. Note that we can set $\theta=0$ without loss of generality because $\theta$ does not have physical meanings [see, Eqs.~(\ref{pot2})-(\ref{pot22})]. First, at the brane $y=0$ the vacuum configuration for $\Phi$ and $\Psi$ is obtained by vanishing of the derivative of $V$ with respect to each component of the scalar fields $\Phi_{j}$ and $\Psi_{j}$. Then, we have six minimization equations for VEVs and six equations for phases. From those equations, we can get~\footnote{Of course, there are trivial solutions $v_{\Phi_{1}}=0, v_{\Psi_{1}}=0$. We have neglected them.}
 \begin{eqnarray}
  \upsilon^{2}_{\Phi}&\equiv&\upsilon^{2}_{\Phi_{1}}=-\frac{18\mu^{2}_{\Phi}+v^{2}_{\Psi}W}
  {2(9\lambda^{\Phi}_{1}+4\lambda^{\Phi}_{3})}
  \neq0~,\qquad\langle\Phi_{2}\rangle=\langle\Phi_{3}\rangle=0~,\nonumber\\
  \upsilon^{2}_{\Psi}&\equiv&\upsilon^{2}_{\Psi_{1}}=-\frac{18\mu^{2}_{\Psi}+v^{2}_{\Phi}W}
  {2(9\lambda^{\Psi}_{1}+4\lambda^{\Psi}_{3})}\neq0~,\qquad\langle\Psi_{2}\rangle=\langle\Psi_{3}\rangle=0~,
  \label{vevPhiPsi}
 \end{eqnarray}
where $W=9\lambda^{\Phi\Psi}_{1}+4\lambda^{\Phi\Psi}_{3}+9\lambda^{\Phi\Psi}_{12}+4\lambda^{\Phi\Psi}_{14}+2(9\lambda^{\Phi\Psi}_{7}+4\lambda^{\Phi\Psi}_{9})\cos2(\gamma-\zeta)$ with $\gamma\equiv\gamma_1,~\zeta\equiv\zeta_1$, and $\upsilon_{\Phi}$ and $\upsilon_{\Psi}$ are real.
With the vacuum alignments of $\Phi,\Psi$ fields, Eq.~(\ref{vevPhiPsi}), minimal condition with respect to $\gamma_{j}, \zeta_{j}$ are given as
 \begin{eqnarray}
  -\frac{\partial V}{\partial \gamma_{1}}\Big|&=&  \frac{\partial V}{\partial \zeta_{1}}\Big|= \frac{v^{2}_{\Phi}v^{2}_{\Psi}}{9}\left(9\lambda^{\Phi\Psi}_{7}+4\lambda^{\Phi\Psi}_{9}\right)\sin2(\gamma-\zeta)=0~,
 \end{eqnarray}
where $\gamma=\gamma_1,~\zeta=\zeta_1$, and $\frac{\partial V}{\partial \gamma_{2,3}}\Big|= \frac{\partial V}{\partial \zeta_{2,3}}\Big|=0$ is automatically satisfied.

Second, at the brane $y=L/2$ the vacuum configuration for $\chi$ and $\eta$ is obtained by vanishing of the derivative of $V$ with respect to each component of the scalar fields $\chi_{j}$ and $\eta$. For simplicity, we consider only the renormalizable terms in $V(\chi)$ and $V(\chi\eta)$. Then, we have seven minimization equations for four VEVs and three phases. From those equations, we can get~\footnote{There are trivial solutions $v_{\chi_{j}}=0$. We have neglected them.}
 \begin{eqnarray}
  \upsilon^{2}_{\chi_{j}}&=&-\frac{m_{\chi}+2(\mu^{2}_{\chi}+v^{2}_{\eta}\lambda^{\chi\eta})\cos2\varphi}
  {12\{(\lambda^{\chi}_{1}+\lambda^{\chi}_{2})\cos4\varphi+\tilde{\lambda^{\chi}_{2}}\cos2\varphi\}}=\upsilon^{2}_{\chi}\neq0~,
  \label{vevchi}
 \end{eqnarray}
where $\upsilon_{\chi}$ is real, and $\varphi_{1}=\varphi_{2}=\varphi_{3}=\varphi$ is used. With the vacuum alignment of $\chi$ fields, Eq.~(\ref{vevchi}), minimal condition with respect to $\varphi_{i}$ is given for $\varphi_{1}=\varphi_{2}=\varphi_{3}$ as
 \begin{eqnarray}
  -\frac{1}{4}\frac{\partial V}{\partial \varphi_{j}}\Big|&=&  v^{2}_{\chi}\left\{v^{2}_{\eta}\lambda^{\chi\eta}+\mu^{2}_{\chi}+3v^{2}_{\chi}\left(\tilde{\lambda}^{\chi}_{2}+4(\lambda^{\chi}_{1}+\lambda^{\chi}_{2})\cos2\varphi_{j}\right)\right\}\sin2\varphi_{j}=0~,
 \end{eqnarray}
with $i=1,2,3$.
And, requiring vanishing of the derivative of $V$ with respect to $\eta$,
 \begin{eqnarray}
  \frac{1}{2}\frac{\partial V}{\partial \eta^{0}}\Big|^{<\chi>=v_{\chi}}_{<\eta^{0}>=v_{\eta}}&=& v_{\eta}\Big\{v^{2}_{\eta}\lambda^{\eta}+\frac{\mu^{2}_{\eta}}{2}+3\lambda^{\Phi\chi}v^{2}_{\chi}\cos2\varphi\Big\}=0~,
  \label{veveta}
 \end{eqnarray}
and, we obtain the VEV of $\eta$ for $\langle\chi\rangle=v_{\chi}e^{i\varphi}(1,1,1)$,
 \begin{eqnarray}
 v^{2}_{\eta}&=&\frac{-\mu^{2}_{\eta}-6v^{2}_{\chi}\lambda^{\chi\eta}\cos2\varphi}{2\lambda^{\eta}}~.
  \label{veveta1}
 \end{eqnarray}
Finally, at the brane $y=L$ the vacuum configuration for $H$ and $G$ is obtained by vanishing of the derivative of $V$ with respect to each component of the scalar fields $H_{i}$ and $G_{i}$. Then, we have eight minimization equations for four VEVs and four phases. From those equations, we can get
 \begin{eqnarray}  \label{vevH}
  \frac{1}{2}\frac{\partial V}{\partial H^{0}_{1}}\Big|^{<G^{0}_j>=v_{G_j}}_{<H^{0}_j>=v_{H_j}}&=& v_{H_1}\Big\{2v^{2}_{H_2}\lambda^{H}_{1}(\cos2\rho_{12}-1)-\lambda^{HG}_{2}v^{2}_{G_2}\Big\}+v_{H_2}\Big\{\mu^{2}_{H}\sin\rho_{12}\nonumber\\
  &-& v_{G_{1}}v_{G_2}\left(2\lambda^{HG}_1\sin\sigma_{12}\sin\rho_{12}-\lambda^{HG}_2\cos(\sigma_{12}+\rho_{12})\right)\Big\}=0~,\nonumber\\
  \frac{1}{2}\frac{\partial V}{\partial H^{0}_{2}}\Big|^{<G^{0}_j>=v_{G_j}}_{<H^{0}_j>=v_{H_j}}&=& v_{H_2}\Big\{2v^{2}_{H_1}\lambda^{H}_{1}(\cos2\rho_{12}-1)-\lambda^{HG}_{2}v^{2}_{G_1}\Big\}+v_{H_1}\Big\{\mu^{2}_{H}\sin\rho_{12}\nonumber\\
  &-& v_{G_{1}}v_{G_2}\left(2\lambda^{HG}_1\sin\sigma_{12}\sin\rho_{12}-\lambda^{HG}_2\cos(\sigma_{12}+\rho_{12})\right)\Big\}=0~,
 \end{eqnarray}
 \begin{eqnarray}
  \frac{1}{2}\frac{\partial V}{\partial G^{0}_{1}}\Big|^{<H^{0}_j>=v_{H_j}}_{<G^{0}_j>=v_{G_j}}&=& v_{G_1}\Big\{2v^{2}_{G_2}\lambda^{G}_{1}(\cos2\sigma_{12}-1)-\lambda^{HG}_{2}v^{2}_{H_2}\Big\}+v_{G_2}\Big\{\mu^{2}_{G}\sin\sigma_{12}\nonumber\\
  &-& v_{H_{1}}v_{H_2}\left(2\lambda^{HG}_1\sin\sigma_{12}\sin\rho_{12}-\lambda^{HG}_2\cos(\sigma_{12}+\rho_{12})\right)\Big\}=0~,\nonumber\\
  \frac{1}{2}\frac{\partial V}{\partial G^{0}_{2}}\Big|^{<H^{0}_j>=v_{H_j}}_{<G^{0}_j>=v_{G_j}}&=& v_{G_2}\Big\{2v^{2}_{G_1}\lambda^{G}_{1}(\cos2\sigma_{12}-1)-\lambda^{HG}_{2}v^{2}_{H_1}\Big\}+v_{G_1}\Big\{\mu^{2}_{G}\sin\sigma_{12}\nonumber\\
  &-& v_{H_{1}}v_{H_2}\left(2\lambda^{HG}_1\sin\sigma_{12}\sin\rho_{12}-\lambda^{HG}_2\cos(\sigma_{12}+\rho_{12})\right)\Big\}=0~.
  \label{vevG}
 \end{eqnarray}
where $\rho_{12}\equiv\rho_1-\rho_2$ and $\sigma_{12}\equiv\sigma_1-\sigma_2$.
And, we obtain the VEVs of $H$ and $G$:
 \begin{eqnarray}
  v_{H_1}&=&\frac{\mu^{2}_{H}\sin\rho_{12}-v_{G_{1}}v_{G_2}Y}{2v_{H_2}\lambda^{H}_{1}(1-\cos2\rho_{12})+\lambda^{HG}_{2}v^{2}_{G_2}/v_{H_2}}~,\nonumber\\
  v_{H_2}&=&\frac{\mu^{2}_{H}\sin\rho_{12}- v_{G_{1}}v_{G_2}Y}{2v_{H_1}\lambda^{H}_{1}(1-\cos2\rho_{12})+\lambda^{HG}_{2}v^{2}_{G_1}/v_{H_1}}~,\nonumber\\
  v_{G_1}&=&\frac{\mu^{2}_{G}\sin\sigma_{12}- v_{H_{1}}v_{H_2}Y}{2v_{G_2}\lambda^{G}_{1}(1-\cos2\sigma_{12})+\lambda^{HG}_{2}v^{2}_{H_2}/v_{G_2}}~,\nonumber\\
  v_{G_2}&=&\frac{\mu^{2}_{G}\sin\sigma_{12}
  -v_{H_{1}}v_{H_2}Y}{2v_{G_1}\lambda^{G}_{1}(\cos2\sigma_{12}-1)-\lambda^{HG}_{2}v^{2}_{H_1}/v_{G_1}}~,
 \label{vevG1}
 \end{eqnarray}
where $Y=2\lambda^{HG}_1\sin\sigma_{12}\sin\rho_{12}-\lambda^{HG}_2\cos(\sigma_{12}+\rho_{12})$.
With the vacuum alignment of $H$ and $G$ fields, Eqs.~(\ref{vevH}) and (\ref{vevG}), minimal condition with respect to $\rho_{j}, \sigma_{j}$ is given as
 \begin{eqnarray}
  -\frac{1}{2}\frac{\partial V}{\partial \rho_{1}}\Big|=\frac{1}{2}\frac{\partial V}{\partial \rho_{2}}\Big|&=&  v_{H_1}v_{H_2}\Big\{-\mu^{2}_{H}\cos\rho_{12}+2v_{H_1}v_{H_2}\lambda^{H}_{1}\sin2\rho_{12}\nonumber\\
  &+& v_{G_1}v_{G_2}\left(2\lambda^{HG}_{1}\sin\sigma_{12}\cos\rho_{12}+\lambda^{HG}_{2}\sin(\sigma_{12}+\rho_{12})\right)\Big\}=0~,\\
  -\frac{1}{2}\frac{\partial V}{\partial \sigma_{1}}\Big|=\frac{1}{2}\frac{\partial V}{\partial \sigma_{2}}\Big|&=&  v_{G_1}v_{G_2}\Big\{-\mu^{2}_{G}\cos\sigma_{12}+2v_{G_1}v_{G_2}\lambda^{G}_{1}\sin2\sigma_{12}\nonumber\\
  &+& v_{H_1}v_{H_2}\left(2\lambda^{HG}_{1}\sin\rho_{12}\cos\sigma_{12}+\lambda^{HG}_{2}\sin(\rho_{12}+\sigma_{12})\right)\Big\}=0~.
 \end{eqnarray}

\section{}
\label{sectionC}
In Eq.~(\ref{lagrangianChLep}) the components $m^{\ell}_{ij}$ are given by
 \begin{eqnarray}
 m^{\ell}_{11}&=&Y^{\mu}_{2}\left(\frac{1-i}{2}\tilde{v}_{H_2}+i\tilde{v}_{H_1}\right)~,\qquad m^{\ell}_{12}=Y^{\mu}_{2}\left(\frac{1-i}{2}\tilde{v}_{H_1}+\tilde{v}_{H_2}\right)~,\nonumber\\
m^{\ell}_{21}&=&Y^{\mu}_{1}\left(\frac{1-i}{2}\tilde{v}_{H_2}+i\tilde{v}_{H_1}\right)~,\qquad m^{\ell}_{22}=Y^{\mu}_{1}\left(\frac{1-i}{2}\tilde{v}_{H_1}+\tilde{v}_{H_2}\right)~.
 \end{eqnarray}
In Eq.~(\ref{lagrangianDown}) the components $m^{d}_{ij}$ are given by
 \begin{eqnarray}
m^{d}_{11}&=&-Y^{d}_{1}\tilde{v}_{H_2}-Y^{a}_{d}\frac{1-i}{4}\tilde{v}_{H_2}+Y^{s}_{d}\left(\frac{2i}{3}\tilde{v}_{H_1}-\frac{1-i}{6}\tilde{v}_{H_2}\right)\nonumber\\
m^{d}_{12}&=&Y^{d}_{1}\tilde{v}_{H_1}+Y^{a}_{d}\left(\frac{1}{2}\tilde{v}_{H_2}-\frac{1-i}{4}\tilde{v}_{H_1}\right)-Y^{s}_{d}\left(\frac{1}{3}\tilde{v}_{H_2}+\frac{1-i}{6}\tilde{v}_{H_1}\right)~,~m^{d}_{13}=\tilde{v}_{\Psi}\frac{2y^{s}_{b}}{3}~,\nonumber\\
m^{d}_{21}&=&-Y^{d}_{1}\tilde{v}_{H_2}+Y^{a}_{d}\left(\frac{1-i}{4}\tilde{v}_{H_2}-\frac{i}{2}\tilde{v}_{H_1}\right)-Y^{s}_{d}\left(\frac{i}{3}\tilde{v}_{H_1}+\frac{1-i}{6}\tilde{v}_{H_2}\right)\nonumber\\
m^{d}_{22}&=&Y^{d}_{1}\tilde{v}_{H_1}+Y^{a}_{d}\frac{1-i}{4}\tilde{v}_{H_1}+Y^{s}_{d}\left(\frac{2}{3}\tilde{v}_{H_2}-\frac{1-i}{6}\tilde{v}_{H_1}\right)~,\quad m^{d}_{23}=-\tilde{v}_{\Psi}\left(\frac{y^{a}_{b}}{2}+\frac{y^{s}_{b}}{3}\right)~,\nonumber\\
m^{d}_{31}&=&-Y^{d}_{1}\tilde{v}_{H_2}+Y^{a}_{d}\frac{i}{2}\tilde{v}_{H_2}+Y^{s}_{d}\left(\frac{1-i}{3}\tilde{v}_{H_2}-\frac{i}{3}\tilde{v}_{H_1}\right)~,\nonumber\\
m^{d}_{32}&=&Y^{d}_{1}\tilde{v}_{H_1}-Y^{a}_{d}\frac{1}{2}\tilde{v}_{H_2}+Y^{s}_{d}\left(\frac{1-i}{3}\tilde{v}_{H_1}-\frac{1}{3}\tilde{v}_{H_2}\right)~,~
 m^{d}_{33}=\tilde{v}_{\Psi}\left(\frac{y^{a}_{b}}{2}-\frac{y^{s}_{b}}{3}\right)~.
 \label{down2}
 \end{eqnarray}
And, in Eq.~(\ref{lagrangianUP}) the components $m^{t}_{ij}$ are given by
 \begin{eqnarray}
m^{t}_{11}&=&Y^{a}_{u}\left(\frac{1-i}{4}\tilde{v}_{G_2}-\frac{i}{2}\tilde{v}_{G_1}\right)-Y^{s}_{u}\left(\frac{1-i}{6}\tilde{v}_{G_2}+\frac{i}{3}\tilde{v}_{G_1}\right)\nonumber\\
m^{t}_{12}&=&Y^{a}_{u}\frac{(1-i)}{4}\tilde{v}_{G_1}+Y^{s}_{u}\left(\frac{2}{3}\tilde{v}_{G_2}-\frac{1-i}{6}\tilde{v}_{G_1}\right)~,\qquad m^{t}_{13}=\tilde{v}_{\Phi}\frac{2y^{s}_{t}}{3}\nonumber\\
m^{t}_{21}&=&Y^{a}_{u}\frac{i}{2}\tilde{v}_{G_1}+Y^{s}_{u}\left(-\frac{i}{3}\tilde{v}_{G_1}+\frac{1-i}{3}\tilde{v}_{G_2}\right)\nonumber\\
m^{t}_{22}&=&-Y^{a}_{u}\frac{1}{2}\tilde{v}_{G_2}+Y^{s}_{u}\left(\frac{1-i}{3}\tilde{v}_{G_1}-\frac{1}{3}\tilde{v}_{G_2}\right)~,\qquad m^{t}_{23}=-\tilde{v}_{\Phi}\left(\frac{y^{a}_{t}}{2}+\frac{y^{s}_{t}}{3}\right)~,\nonumber\\
m^{t}_{31}&=&-Y^{a}_{u}\frac{1-i}{4}\tilde{v}_{G_2}+Y^{s}_{u}\left(\frac{2i}{3}\tilde{v}_{G_1}-\frac{1-i}{6}\tilde{v}_{G_2}\right)~,\nonumber\\
m^{t}_{32}&=&Y^{a}_{u}\left(\frac{1}{2}\tilde{v}_{G_2}-\frac{1-i}{4}\tilde{v}_{G_1}\right)-Y^{s}_{u}\left(\frac{1-i}{6}\tilde{v}_{G_1}+\frac{1}{3}\tilde{v}_{G_2}\right)~,\quad m^{t}_{33}=\tilde{v}_{\Phi}\left(\frac{y^{a}_{t}}{2}-\frac{y^{s}_{t}}{3}\right)~.
 \label{up1}
 \end{eqnarray}

\acknowledgments{
We thank prof. E.~J.~Chun for useful discussions.}


\end{document}